\documentclass[twocolumn]{aa} % for a paper on 1 column  
\DeclareTextSymbol{\degre}{OT1}{23}

\usepackage{graphicx,natbib}
\usepackage{txfonts}
\begin{document}

\title{From planetesimals to planets: volatile molecules}

\authorrunning{Marboeuf et al.}

\author{Ulysse~Marboeuf\inst{1}, Amaury~Thiabaud \inst{1}, Yann Alibert \inst{1, 2}, Nahuel Cabral\inst{1}, \& Willy Benz \inst{1}}
\institute{${}^1$Physics Institute and Center for Space and Hability, University of Bern, Bern, Switzerland\\
${}^2$ Observatoire de Besan\c con, France\\
\email{ulysse.marboeuf@space.unibe.ch}}

\date{Received ??; accepted ??}

% \abstract{}{}{}{}{} 
% 5 {} token are mandatory
  
  \abstract
% {} leave it empty if necessary  sta
{Solar and extrasolar planets are the subject of numerous studies aiming to determine their chemical composition and internal structure. In the case of extrasolar planets, the composition is important as it partly governs their potential habitability. Moreover, observational determination of chemical composition of planetary atmospheres are becoming available, especially for transiting planets.
}
{The present works aims at determining the chemical composition of planets formed in stellar systems of solar chemical composition. The main objective of this work is to provide valuable theoretical data for models of planet formation and evolution, and future interpretation of chemical composition of solar and extrasolar planets.}
{We have developed a model that computes the composition of ices in planets in different stellar systems with the use of models of ice and planetary formation.}
{We provide the chemical composition, ice/rock mass ratio and C:O molar ratio for planets in stellar systems of solar chemical composition. From an initial homogeneous composition of the nebula, we produce a wide variety of planetary chemical compositions as a function of the mass of the disk and distance to the star. 
%Planets formed in stellar systems are mainly rocky (20\%), icy/ocean (66\%) or gas giant planets (14\%). 
The volatile species incorporated in planets are mainly composed of H$_2$O, CO, CO$_2$, CH$_3$OH, and NH$_3$. Icy or ocean planets have systematically higher values of molecular abundances compared to giant and rocky planets.
 Gas giant planets are depleted in highly volatile molecules such as CH$_4$, CO, and N$_2$ compared to icy or ocean planets. The ice/rock mass ratio in icy or ocean and gas giant planets is, respectively, equal at maximum to 1.01$\pm$0.33 and 0.8$\pm$0.5, and is different from the usual assumptions made in planet formation models, which suggested this ratio to be 2-3. The C:O molar ratio in the atmosphere of gas giant planets is depleted by at least 30\% compared to solar value.}
{}


\maketitle

\section{Introduction}
The determination of the chemical composition of planets has been the subject of numerous studies. 
The study of the composition of volatile molecules incorporated in planets is an important parameter in planetary internal structure and thermal evolution models of these bodies. 
Numerous studies focused on the internal structure and chemical composition (mainly atmospheres or ices) of solar (Gautier et al. 2001a,b; Hersant et al. 2004) and extrasolar planets (Hersant et al. 2004; Marboeuf et al. 2008; Bond et al. 2010; Mousis et al. 2010; Johnson et al. 2012). 
Since the discovery of the first extrasolar planet in 1995, more than 1000 have been discovered by radial velocity measurements, microlensing experiments, and photometric transit observations (Baraffe et al. 2010; Lunine 2011; Griffith et al. 2011). 
These exoplanets cover a wide range of masses and sizes, among which exoplanets of sub-Mercury sizes (Kepler 37c, Barclay et al. 2013) or a few tens of Jupiter masses (see Udry \& Santos 2007 for a review).
Some molecular species such as H$_2$O, CO, CO$_2$, and CH$_4$ have already been detected in the troposphere and/or stratosphere of extrasolar giant planets (Beaulieu et al. 2008, 2010; Barman 2007, 2008; Burrows et al. 2005, 2007, 2008; Charbonneau et al. 2002, 2005, 2008; D\'esert et al. 2009; Fortney \& Marley 2007; Griffith et al. 2011; Grillmair et al. 2008; Lee et al. 2012; Madhusudhan \& Seager 2009, 2010, 2011;  Madhusudhan et al. 2011; Madhusudhan 2012; Snellen et al. 2008; Swain et al. 2008, 2009a,b; Tinetti et al. 2007, 2010, 2012; Waldmann et al. 2012). Some numerical studies have also been developed to theoretically predict and explain these observations (e.g., Marboeuf et al. 2008; Johnson et al. 2012) and the implication of this composition on the planetary spectra (Moses et al. 2013) that could be observed from Earth. Moreover, the Exoplanet Characterisation Observatory (EChO), will be the first dedicated mission to investigate the physics and chemistry of exoplanetary atmospheres in order to characterize the physical conditions of their formation and evolution.
However, none of these studies has computed the planetary formation in a self-consistent way.

We have recently extended our planet formation model to the formation of planetary systems (see Fortier et al. 2013, Alibert et al. 2013), and our aim, in this paper and a companion one (Thiabaud et al. 2014, hereafter paper 1), is to determine the composition of planet, in a way that is self-consistent with the formation process as modeled in our approach. We have first computed the composition of planetesimals for a population of protoplanetary disk (see Marboeuf et al. submitted, hereafter paper 2), and at any distance to the central star, considering both the icy species (topic of this paper), and refractory species (see paper 1).
In this paper, we combine the results of the formation model (which provides the amount of planetesimals accreted by every planet as a function of the distance to the central star) with the composition model presented in the paper 2, to derive the final composition of planets. The main objective of this work is therefore to provide valuable theoretical data for models of planet formation and evolution, and future interpretation of chemical composition of both solar and extrasolar planets. 

This article is organized as follows. In a first step (Sect.~\ref{models}), we present the planetary formation model. In a second step (Sect.~\ref{parameter}), we present the physical assumption and parameters adopted. We then calculate the chemical composition of planets formed in the protoplanetary disk (Sect.~\ref{results_planets}), and discuss the results (Sect.~\ref{discussion}). The last section (Sect.~\ref{conclusion}) is devoted to conclusions.

\section{Description of the physical models used in this work \label{models}}

In order to determine the chemical composition of planets we consider several planetary systems assumed to emerge from a protoplantary disk whose initial density profile, mass, and lifetime is different, and follows, as closely as possible, observational characteristics (see Mordasini et al. 2009, Alibert et al. 2013). We assume, when computing the formation of planets, that planetesimals are already formed at the beginning of the planetary system formation process (initial time of the model) by a process that is presently still a matter of research. 
%debate.
Planetesimals are believed to form from the coagulation of small grains (icy and/or rocky), whose chemical composition is itself provided by the condensation sequence of gas in the protoplanetary disks (see paper 2).
This provides us with the composition of grains, which is also assumed to be identical to that of planetesimals formed at the same location in the disk.
This approach is obviously simplified, since the chemical composition of planetesimals results from the condensation/clathrate sequence at the location of the grains (see paper 2), and we do not include the radial drift of small icy grains. We then use our planetary formation model (Alibert et al. 2005, Mordasini et al. 2009, Mordasini et al. 2012a,b, Fortier et al. 2013, Alibert et al. 2013) to compute the chemical composition of planets formed in the disk.

\subsection{Model of dynamical planet migration and calculation of the chemical composition of planets \label{model_accretion}}

The initial locations of planetary embryos (ten embryos are considered in the models presented in this paper) are selected at random following a uniform distribution in log. The composition of the planetary embryos (whose mass is on the order of lunar mass - $10^{-2} M_\oplus$) is assumed to be the same as planetesimals condensed at its location (see paper 2). In other words, we assume that the initial planetary embryos are formed locally out of planetesimals formed at the same location.

The model of planet formation allows us to determine the formation path of the ten initial embryos in the protoplanetary disk and their growth during their migration for several disks (different initial surface densities $\Sigma_0~(T_0,P_0,r)$\footnote{$\Sigma_0$ is the initial surface density at 5.2 AU.} of the disk), and for different initial distances to the star.
The mass of planetesimals accreted is given by

\begin{equation}
\frac{d M_p(r)}{dt} = \pi R^2_c \Sigma_p \Omega F_g   \qquad (kg \, s^{-1}),
\end{equation}
where $\Sigma_p$ (g.cm$^{-2}$) is the surface density of solids in feeding zone, $R_c$ (AU) the capture radius of the planetesimal, $F_g$ the focusing factor, and $\Omega$ the Keplerian frequency

\begin{equation}
\Omega^2 = \frac{G M_*}{r^3}    \qquad (s^{-2})
\end{equation}
with $r$ the distance to the star, $G$ the gravitational constant, and $M_*$ the mass of the central star, assumed to be solar in this study.

\begin{figure}[h]
\begin{center}
\includegraphics[width=5.cm]{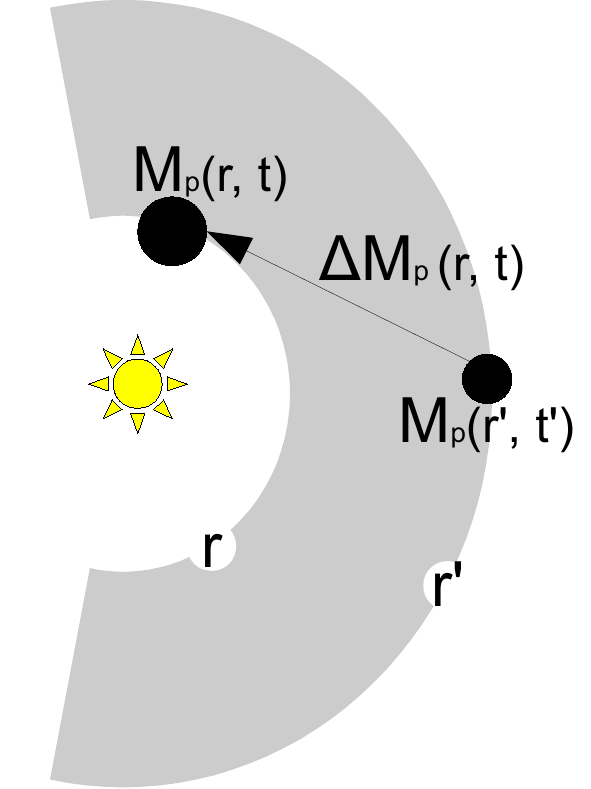}
\caption{Schematic view of the migration and grow of a protoplanet by accretion of planetesimals in the protoplanetary disk between positions to the star $r$ and $r'$.}
\label{figp1}
\end{center}
\end{figure}

During the migration of the planet in the protoplanetary disk, from a distance $r'$ to a distance $r$ from the star (see fig.~\ref{figp1}), the variation of the mass $\Delta m_x^p(r)$ of the species $x$ in the planet $p$ after accretion of planetesimals is defined as
\begin{equation}
\Delta m_x^p(r, t) =\Gamma_x^g(r) \Delta M_p(r, t)   \qquad (kg),
\end{equation}
where $\Delta M_p(r,t)$ is the mass of planetesimals accreted by the planet $p$ at time $t$ and $\Gamma_x^g(r)$ the chemical abundance in planetesimals of species $x$.

At each time step, we compute the mass $M_p(r, t)$ of the planet $p$ and the mass $m_x^p(r, t)$ of volatile molecules $x$ in the planet $p$ after accretion of new material $\Delta M_p(r, t)$ at the position $r$ and time $t$,
\begin{equation}
M_p(r, t) = M_p(r', t') + \Delta M_p(r, t)   \qquad (kg),
\end{equation}
where $M_p(r')$ is the mass of the planet at the distance $r'$ and time $t'$ from the star before accretion of planetesimals and

\begin{equation}
m_x^p(r, t) = m_x^p(r', t') + \Delta m_x^p(r, t)   \qquad (kg)
\end{equation}
with $m_x^p(r', t')$ the mass of volatile molecules $x$ in the planet $p$ at time $t'$ before accretion of planetesimals.
At the end of the computation, the molar ratio of volatile molecule $x$ relative to all volatile molecules $\Gamma_x^p(r)$ is then given by the relation
\begin{equation}
\Gamma_x^p(r) = \frac{m_x^p(r)}{\sum m_x^p(r)}.
\end{equation}

The process of accretion of planetesimals onto embryos is assumed to retain the original mass of species initially incorporated in icy bodies. The abundance of volatile molecules initially contained in planetesimals does not undergo chemical changes during the melting, the sublimation, and break up of bodies onto planets. Moreover, the migration of planets toward the star which could induce chemical reactions due to an increase in the surface temperature and temperature-pressure of planetary atmospheres is assumed not to change the abundances of species in planets.

\section[]{Physical assumptions and parameters adopted in models \label{parameter}}

In this paper, we focus on the same volatile components used in the study of planetesimals (see paper 2): H$_2$O, CO$_2$, CO, CH$_3$OH, CH$_4$, NH$_3$, N$_2$, and H$_2$S. These volatile components are the major abundant species observed in comets (Bockel{\'e}e-Morvan et al. 2004; Mumma \& Charnley 2011) and planets. We assume that all volatile molecules are composed of H, O, C, N, and S atoms in solar abundances (Lodders 2003). The chemical composition of planets is determined by taking into account both refractory and volatile components. Refractory elements include both minerals and refractory organics (see paper 1 for details). 
Based on the study on the chemical composition of planetesimals, we study several extremes cases by varying parameters based on the formation of clathrates and refractory organics before the condensation of volatile species at lower temperature, the abundance of CO (rich CO and poor CO models assuming CO:CO$_2$=5:1 and 1:1, respectively), and the radiation through the disk.
As shown in paper 2, the formation of clathrates changes the chemical composition of planetesimals by increasing the abundance of CO, CH$_4$, H$_2$S, and N$_2$. The formation of refractory organics before the condensation of ices induces a lower amount of ices of several tens of \% in planetesimals compared to models without organics.
The abundance of CO in planetesimals decreases in parallel to that in the disk.
The irradiation of the disk (hereafter irradiated models) shifts the ice line positions of all volatile species by about 0.5-1 AU farther in the disk and reduces the amount of highly volatile species incorporated in planetesimals such as CH$_4$, CO, and N$_2$, compared to non-irradiated models. As discussed in paper 2, the irradiated and non-irradiated models do not properly model the physics in the disk, but allow us to frame the thermodynamic conditions of ice formation in the cooling disk and migration of planets (see paper 1) respectively near and far away from the star.   
The chemical processes which could change abundances of volatile species in warm surface layers of the protoplanetary disk are not taken into account in this study. All the volatile molecules (H$_2$O, CO$_2$, CH$_3$OH, CH$_4$, NH$_3$, and H$_2$S), except CO and N$_2$, are easily destroyed in the gas phase of the surface layers of the disk (see Visser et al. 2011) by reactions with UV and X-ray photons from the central star as well as UV photons and cosmic rays originating from the interstellar medium. They do not keep high abundances once they are evaporated from grains (Doty et al. 2002; Nomura \& Millar 2004). However, grain surface reactions in the disk will recover the loss, and the abundances of volatile molecules in ice may not be very different from those adopted in planetesimals for this study (see paper 2 for more explanation).  

Finally, we analyze the chemical composition of planets formed in the different synthesis planetary systems presented in Alibert et al. (2013) using 16 different initial chemical compositions of planetesimals (see paper 2) and using different initial surface densities at 5.2 AU (varying from 3 g.cm$^{-2}$ to about 700 g.cm$^{-2}$), density slopes (varying from 0.4 to 1.1), and characteristic scaling radius (varying from 14 AU to 198 AU) of the disk (see Sect. 3.3 in paper 1, and Alibert et al. 2013).
The different initial physical conditions of disks induce different initial thermodynamic conditions of ice formation (see paper 2) and planetary evolution as shown in Fig.~7 of paper 2.
As mentioned in Sect.~\ref{models}, each simulation starts with ten planetary embryos that grow and interact within the same protoplanetary disk. The characteristics of the protoplanetary disk (namely mass, lifetime, and gas-to-solids ratio) are varied from one simulation to the other in order to explore the different possible initial conditions. Depending on the characteristics of the simulation, the composition of planetesimal will change, as well as the parts of the disk sampled by the different planets, providing a diversity of composition.

At the beginning of the simulations, the protoplanets have a lunar mass (mass of about 10$^{-2}$ M$_\oplus$) and the chemical composition of the planetesimals formed at that location. With time, protoplanets migrate and grow (see Sect.~\ref{model_accretion}) by accretion of planetesimals whose chemical composition is described in paper 2. As a result, they are enriched or depleted in volatile molecules compared to initial protoplanets. The mass of planetesimals accreted by a planet depends on the solid surface density and the protoplanet's efficiency in capturing solid planetesimals. 
Unless stated explicitly, figures described in Sect.~\ref{results_planets} present the mass of solid components (ices, rocks, and refractory organics) M$^s_{pl}$ in planets 	and do not include the mass of the gas. This means that we do not include in our model the possibility of capture of volatile species as part as the accreted gas (which is therefore assumed to contain only H and He).

\section{Chemical composition of planets \label{results_planets}}

\subsection{Type of planets and Ice/rock mass ratio}
\begin{figure*}
\begin{center}
\includegraphics[width=14.cm]{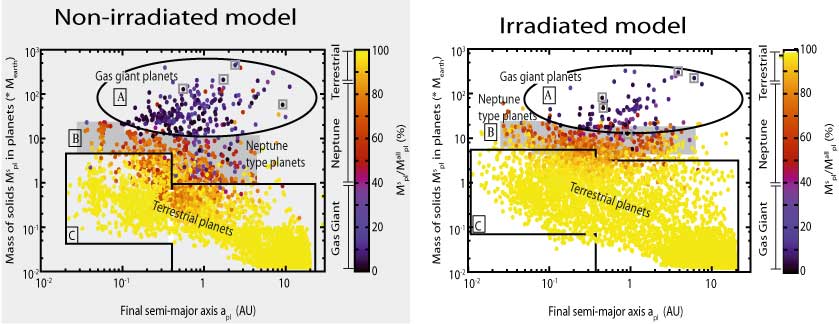}
\caption{Mass ratio of solids M$^s_{pl}$ (ices, rocks, and refractory organics) relative to the total planetary mass M$^{all}_{pl}$ as a function of their final semi-major axis $a_{pl}$ and solid mass M$^s_{pl}$, for non-irradiated (left panel) and irradiated (right panel) models. The light gray boxes are stars with total masses greater than  13 Jovian mass which could be brown dwarfs.}
\label{figp2}
\end{center}
\end{figure*}

\begin{figure*}
\begin{center}
\includegraphics[width=15.cm, angle=0]{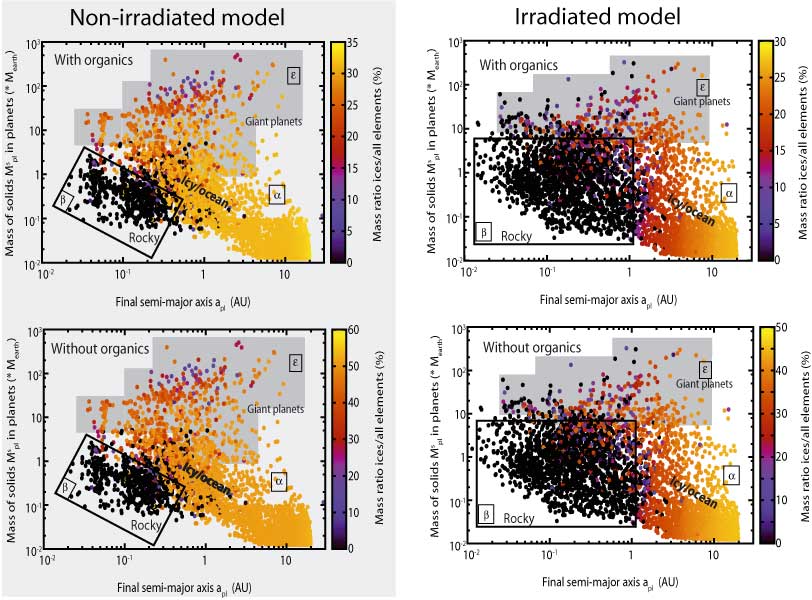}
\caption{Average mass ratio of volatile molecules relative to the total solid mass of planets as a function of the final semi-major axis a$_{pl}$ and the mass of the solid component M$^s_{pl}$, for non-irradiated (left panel) and irradiated (right panel) models, with (upper panel) and without (lower panel) refractory organics.}
\label{figp3}
\end{center}
\end{figure*}

\begin{figure}
\begin{center}
\includegraphics[width=9.cm, angle=0]{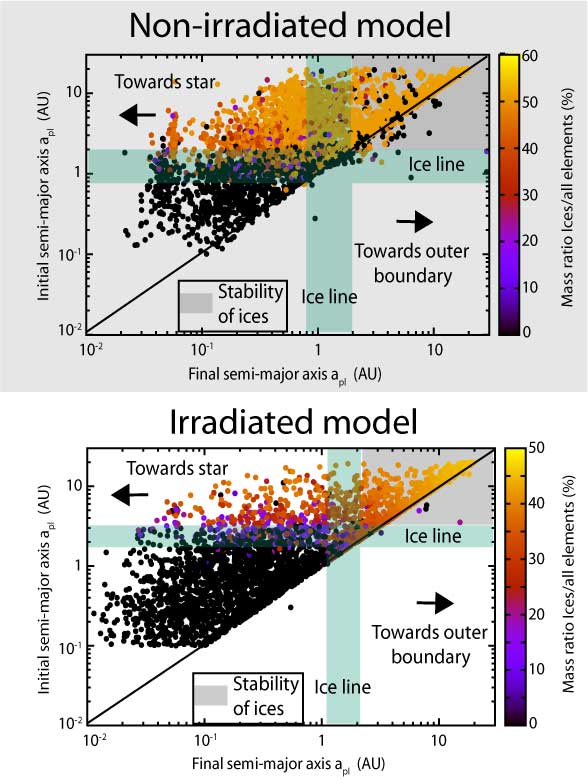}
\caption{Mass ratio of volatile molecules relative to the mass of solids in the planets, as a function of initial and final semi-major axis a$_{pl}$, for non-irradiated (upper panel) and irradiated (lower panel) models. The blue area corresponds to the ice water line position for most simulations. Gray area corresponds to the zone of stability of ices beyond the ice line. Models without refractory organics assume CO:CO$_2$ = 1:1.}
\label{figp6}
\end{center}
\end{figure}

\paragraph{Mass fraction of the solid component of planets}
Figure~\ref{figp2} shows the mass ratio of solid components (core of the planet) M$^s_{pl}$ relative to the total mass of the planet (M$^{all}_{pl}$, including solid components and envelope) in planets as a function of their final semi-major axis a$_{pl}$ and of the mass of the solid components M$^s_{pl}$\footnote{The term "solid component" does not necessarily mean that these elements remain in the solid phase (the planetary core), but only that they have been captured under the form of planetesimals. Similarly, the term "mass of gas" refers to the mass that has been accreted under the form of gas, including possible small dust.}, for non-irradiated (left panel) and irradiated (right panel) models. 
For both models, the population of planets are similar and can be divided into three groups. The first (hereafter group $A$) corresponds to gaseous giant planets with M$^s_{pl}$ $\geq$ 20 M$_\oplus$ for both non-irradiated and irradiated models. The mass of the gas envelope is at least 60 wt\% of the total mass M$^{all}_{pl}$ of planets (black and blue dots).
The second group (gray area, hereafter group $B$) includes less massive planets, with a mass in the range of $\approx$ 1-5 M$_\oplus$ $\leq$ M$^s_{pl}$ $\leq$ 10-20 M$_\oplus$ (resp. $\approx$ 5 M$_\oplus$ $\leq$ M$^s_{pl}$ $\leq$ $\approx$ 20 M$_\oplus$) for  non-irradiated (resp. irradiated) models. 
The mass of solids M$^s_{pl}$ included in these planets varies from about 40 wt\% to about 90 wt\% of the total mass M$^{all}_{pl}$ of planets (purple and orange dots). 
The last group (hereafter group $C$) includes low mass planets, with M$^s_{pl}$ $\leq$ 1-5 M$_\oplus$ (resp. M$^s_{pl}$ $\leq$ 5-10 M$_\oplus$) for  non-irradiated (resp. irradiated) models and a very small mass fraction of gas ($\leq$ 10\%, yellow dots). 
Some planets (light gray boxes, 4 in non-irradiated and 5 in irradiated models) have total masses greater than 13 Jovian mass and could be brown dwarfs rather than gas giant planets.
As can be seen in figure~\ref{figp2}, the mass of gas increases with the mass of heavy elements, a result typical of planets formed in the core-accretion model. 
The major change with the irradiated models is the number of low and high mass planets. The number of low/high mass planets increases/decreases slightly compared to the non-irradiated models. As shown below (see Fig.~\ref{figp3}), this is due to the depletion of ices in the disk and to the change of the migration pathway of planets between non-irradiated and irradiated models (see Fig.~\ref{figp6}).

We note that the maximum mass of heavy elements in gaseous giant planets, which reaches about 500 M$_{earth}$, is probably favored by the hypothesis that all planetesimals have the same radius (on the order of 100 m) and this radius does not evolve with time (see Fortier et al. 2013; Alibert et al. 2013). A size distribution of planetesimals and a radius evolution would probably change the maximum mass of planets formed.

\paragraph{Fraction of ices relative to the total solid mass} Figure \ref{figp3} shows the average mass ratio of volatile molecules, relative to the total mass of heavy elements (volatile molecules, and rocks (including minerals and organics compounds), excluding the gaseous envelope) as a function of the mass of solid components M$^s_{pl}$ and the semi-major axis a$_{pl}$ of planets at the end of the computation, for non-irradiated and irradiated models, both with and without refractory organics, and taking into account all chemical variations (CO:CO$_2$ variation and clathrate formation).
Whatever the model used, the planets are chemically divided into three parts and cover approximately the subdivisions $A$, $B$, and $C$ described in Fig.~\ref{figp2}: 

$\bullet$ 1) The first area (hereafter area $\alpha$) is composed of planets with a solid mass M$^s_{pl}$ ranging from 0.01 M$_\oplus$ to 5 M$_\oplus$ (resp. 0.01 M$_\oplus$ to about 10 M$_\oplus$) for non-irradiated (resp. irradiated) models. These planets are mainly icy or ocean planets (hereafter icy/ocean planets) with low mass and a very small gaseous envelope (see Fig.~\ref{figp2}). 
They are characterized by a final semi-major axis a$_{pl}$ ranging from 0.4 AU to 20 AU (resp. 1 AU to 20 AU) for non-irradiated (resp. irradiated) models. We note that some icy/ocean planets are outside of the ranges described; some planets are located near the star between 0.04 AU and 0.4 AU (0.1 AU and 1 AU) for non-irradiated (resp. irradiated) models and could be ocean planets.
Typically, the solid component of icy/ocean planets is composed of 31$\pm$8 wt\% (resp. 50$\pm$12 wt\%) of ices in non-irradiated models with (resp. without) refractory organics. Variations represent the standard deviation 2$\sigma$ of values, taking into account all chemical changes (formation of clathrates, CO:CO$_2$ variation). Planets in irradiated models (without refractory organics compounds) present lower ratios with a radial gradient of the amount of ices.
Average values ($\pm$ 2$\sigma$) of the fraction of ices in planets are 23$\pm$10 wt\% (resp. 40$\pm$14 wt\%) in icy/ocean planets, with (resp. without) refractory organics. This leads to icy/ocean planets with average ice/rock mass ratio varying from 0.30$\pm$0.16 (irradiated models with refractory organics) to 1.01$\pm$0.33 (non-irradiated models without refractory organics). Such values frame the ones of some icy satellites and KBOs in the solar system: densities of Jovian icy satellites Ganymede and Callisto; Saturn's moon Titan; Neptune's moon Triton; and KBOs Quaoar, Orcus, Eris, and the Pluto-Charon binary ($\approx$ 1.8-2.3 g.cm$^{-3}$) suggest comparable masses of rock and ices (Showman \& Malhotra 1999; Buie et al. 2006; Person et al. 2006; Brown \& Schaller 2007; McKinnon et al. 2008; Sotin et al. 2009; Lunine et al. 2010).

$\bullet$ 2) The second area (hereafter area $\beta$) contains dry terrestrial planets (little or no H$_2$ and He gas) with at maximum 1-2 wt\% of volatile molecules (mainly H$_2$O) for all models. This area is largely composed of rocky planets that represent approximately 20\% (resp. $\approx$ 40\%) of all the planets for non-irradiated (resp. irradiated) models. The solid mass M$^s_{pl}$ and positions a$_{pl}$ of these planets vary from 0.5 M$_\oplus$ to about 2 M$_\oplus$ (resp. in the range 0.3-10 M$_\oplus$), and from 0.02 AU to 0.5 AU (resp. from 0.01 AU to 1.5 AU), for non-irradiated models (resp. irradiated models).
We note that there are some rocky planets outside of this area with masses varying from 0.5 M$_\oplus$ to 5 M$_\oplus$ (resp. 15 M$_\oplus$) and positions varying from 0.05 AU to 3 AU (resp. 3 AU) for non-irradiated (resp. irradiated) models.

$\bullet$ 3) The third area (hereafter area $\epsilon$) is composed of gas giant and Neptune-type planets (hereafter called giant planets) with a solid mass M$^s_{pl}$ ranging from 1-5 M$_\oplus$ to 500 M$_\oplus$ and a semi-major axis a$_{pl}$ varying from 0.03 AU to 10 AU, for both irradiated and non-irradiated models. 
The average mass fraction of volatile molecules (relative to the total solid mass) for planets in this area is 26$\pm$11 wt\% (resp. 43$\pm$17 wt\%) for non-irradiated models with (resp. without) refractory organics. For irradiated models, this fraction decreases and represents about 16$\pm$13 wt\% (resp. 28$\pm$23 wt\%) with (resp. without) refractory organics. This leads to planets with average ice/rock mass ratios varying from 0.19$\pm$0.2 (irradiated models with refractory organics) to 0.8$\pm$0.46 (non-irradiated models without refractory organics).

In summary, we obtain for both irradiated and non-irradiated models, three types of planets: giant planets (1-5 $\leq$ M$^s_{pl}$ $\leq$ 500 M$_\oplus$), rocky planets (M$^s_{pl}$ $\leq$ 2-5 M$_\oplus$, and a$_{pl}$ $\leq$ 0.01-2 AU), and icy/ocean planets (M$^s_{pl}$ $\leq$ 5-10 M$_\oplus$, and a$_{pl}$ $\geq$ 0.04 AU). The mass, position relative to the star, and the amount of ices in planets are mainly functions of the surface density of the disk, the formation of refractory organics, and the radiation in the disk. The full irradiation of the disk shifts the ice line positions (see paper 2) and decreases the amount of ices incorporated in planets by up to 5-15 wt\% compared to non-irradiated dics. Moreover, the change of the thermodynamic conditions in the disk for irradiated models changes the pathway of planets which migrate more to the star (see Fig.~\ref{figp6} which shows the mass ratio of ices relative to the total solid mass of planets, models without organics, for the non-irradiated and irradiated models as a function of their initial and final semi-major axis a$_{pl}$) and incorporate more rocks than ices. 
This results in a greater number of less massive and rocky planets near the star and fewer gas giant planets in the disks.
We also note that whatever the models, planets migrate mainly towards the star while only a few migrate outwards.

\begin{figure}
\begin{center}
\includegraphics[width=9.cm]{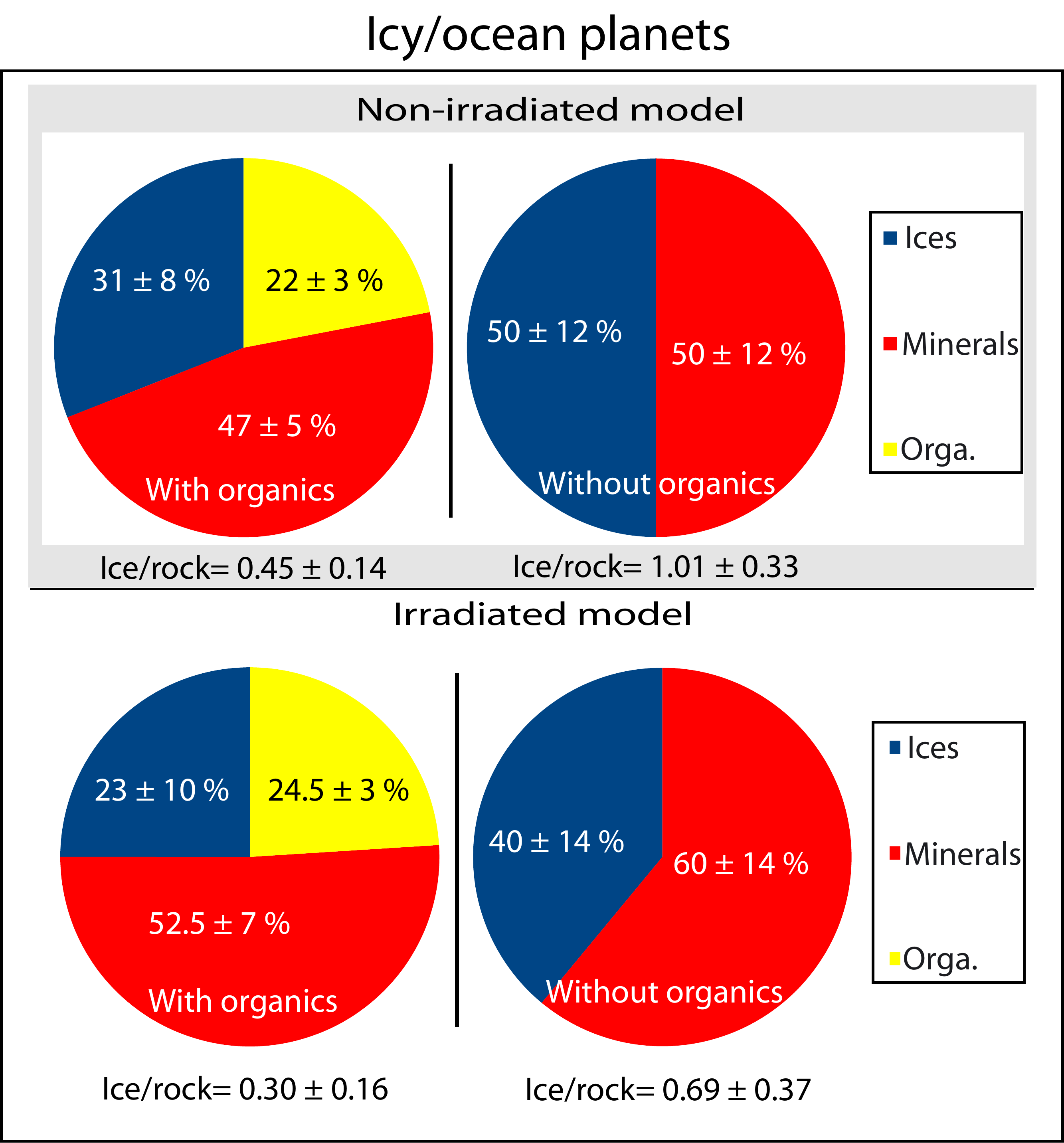}
\caption{Average (with standard deviation 2$\sigma$) mass ratio of ices, minerals, and refractory organics in icy/ocean planets for all models, taking into account chemical variations(CO:CO$_2$ variation and clathrate formation).}
\label{figp4}
\end{center}
\end{figure}

\begin{figure}
\begin{center}
\includegraphics[width=9.cm]{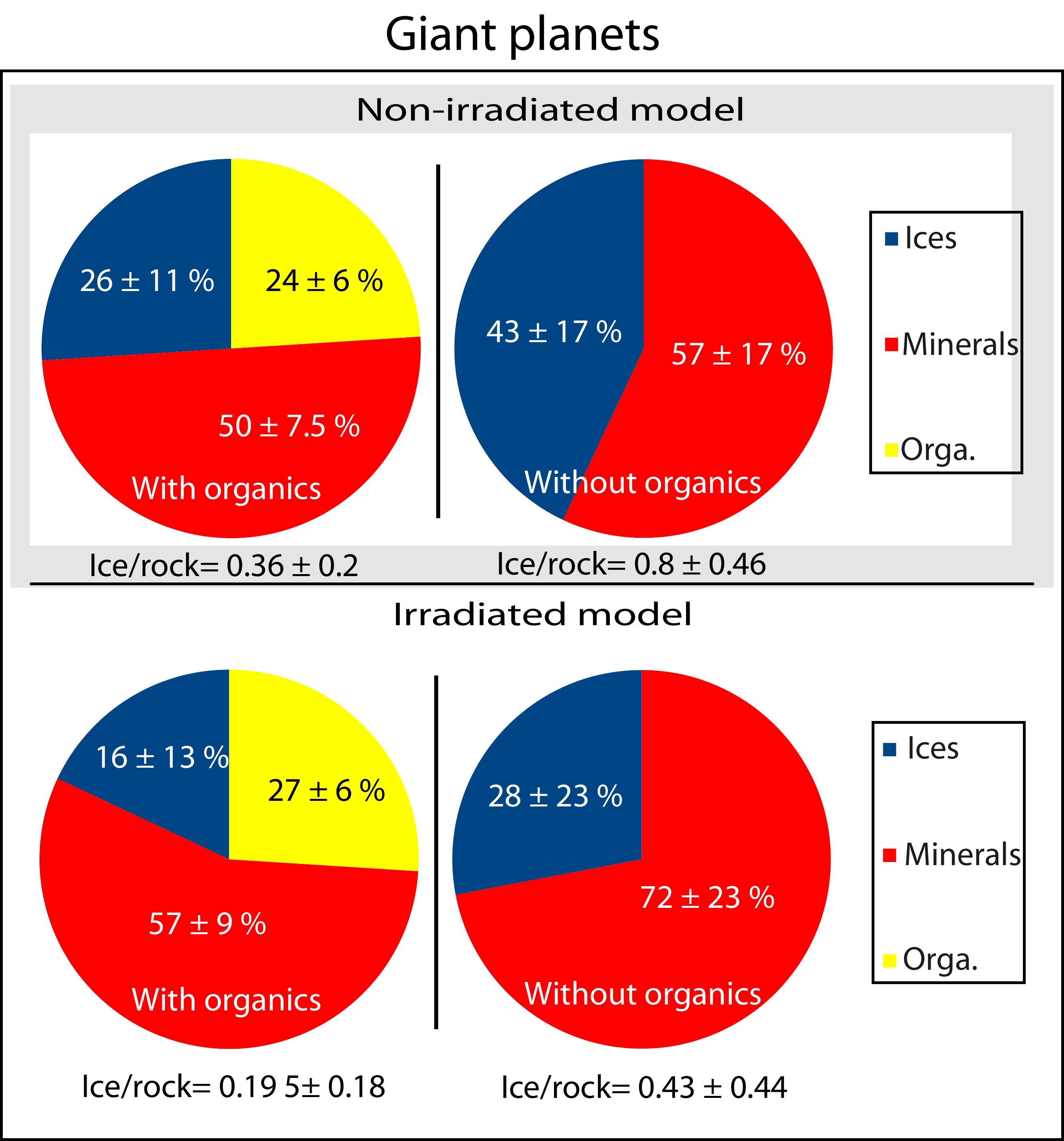}
\caption{Same as Fig.~\ref{figp4} but for giant planets.}
\label{figp5}
\end{center}
\end{figure}

Figures \ref{figp4} and \ref{figp5} present the average mass ratio of ices, minerals, and refractory organics relative to all solid elements, and the ice/rock mass ratio (both with standard deviation $\pm$2$\sigma$) in icy/ocean and giant planets, respectively, and for all models, taking into account the formation of clathrates and the variation of the abundance of CO in planetesimals.

\subsection{Abundances of volatile molecules and atoms in planets}
\begin{figure*}
\begin{center}
\includegraphics[width=15.cm]{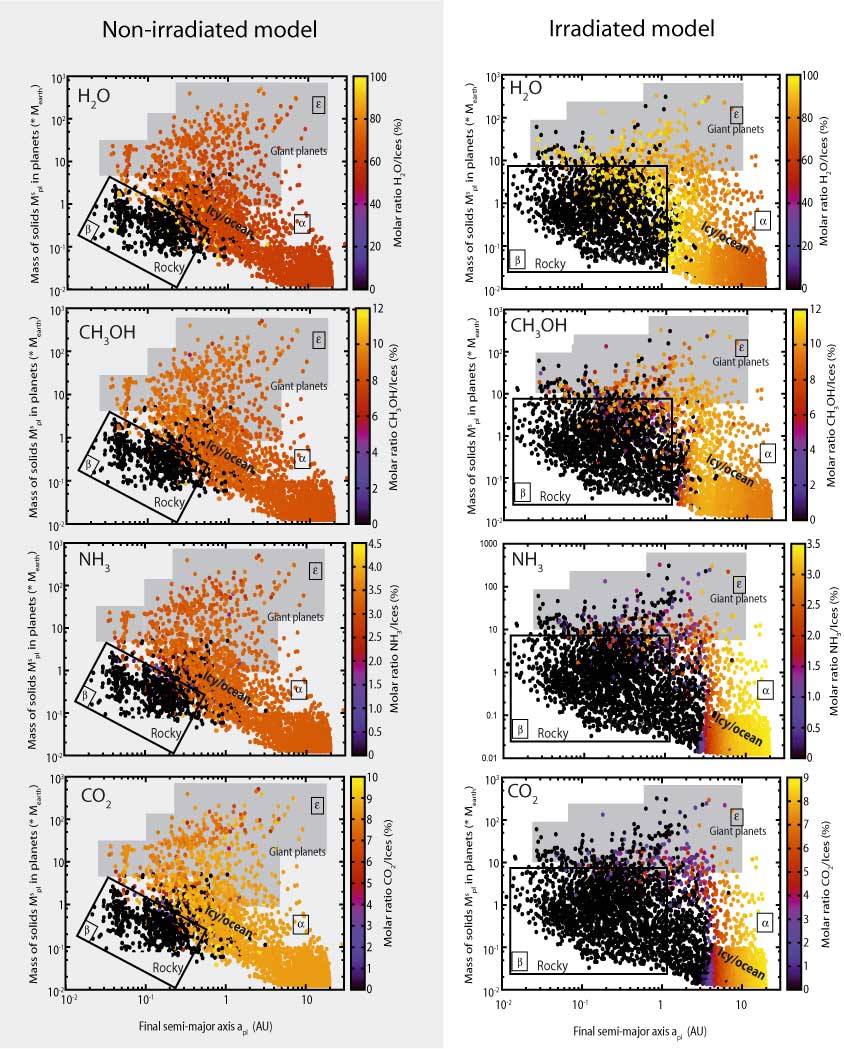}
\caption{Average molar ratio of species H$_2$O, CH$_3$OH, NH$_3$, and CO$_2$ relative to all volatile species in planets as a function of the solid mass M$^s_{pl}$ and final semi-major axis a$_{pl}$ for non-irradiated and irradiated models, taking into account all chemical changes (CO:CO$_2$ variation and clathrate formation).}
\label{figp8}
\end{center}
\end{figure*}

\begin{figure*}
\begin{center}
\includegraphics[width=15.cm]{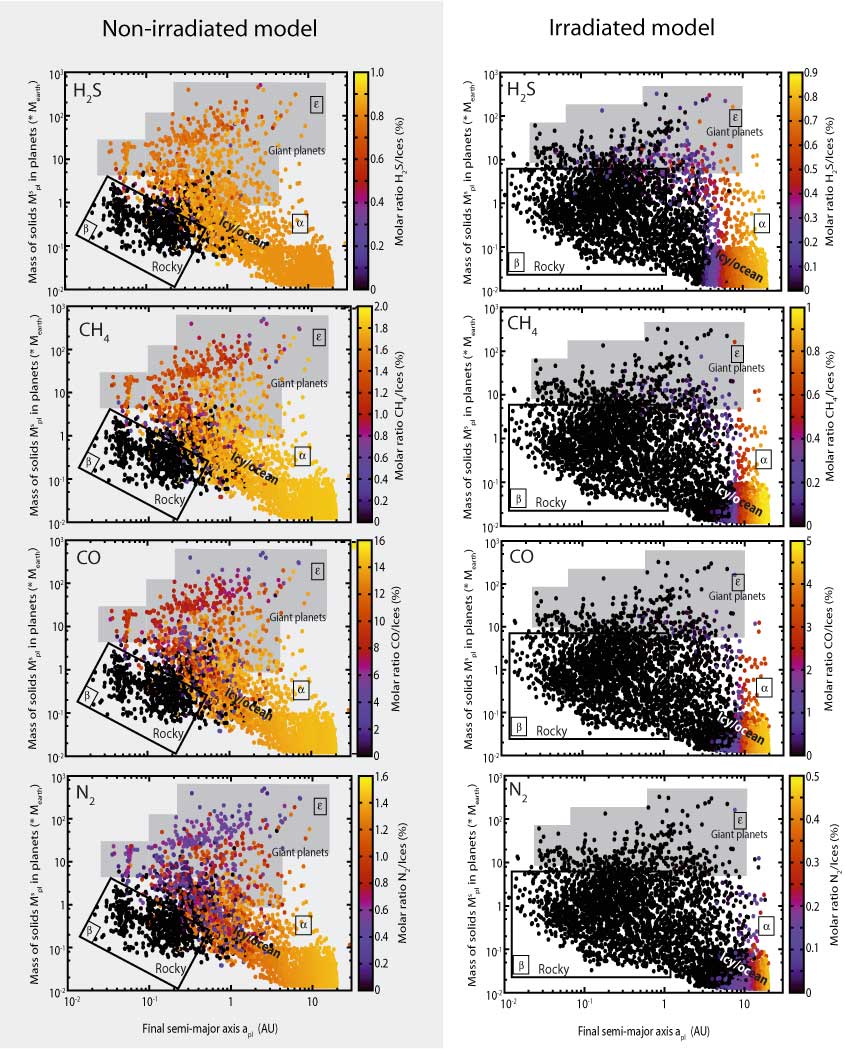}
\caption{Same as Fig.~\ref{figp8} but for species H$_2$S, CH$_4$, CO, and N$_2$.}
\label{figp9}
\end{center}
\end{figure*}

\begin{figure}
\begin{center}
\includegraphics[width=9.cm]{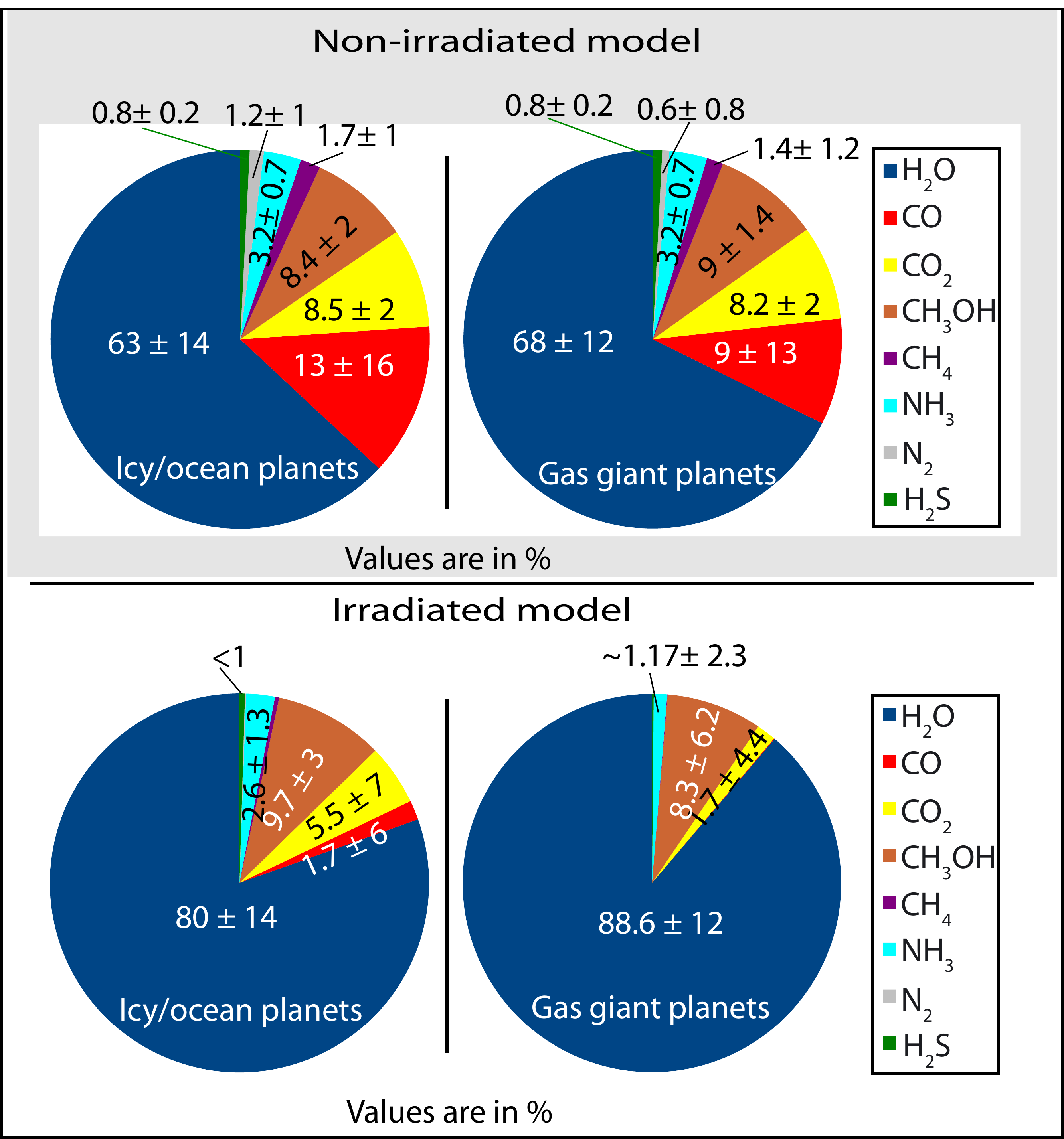}
\caption{Average molar ratio (with standard deviation 2$\sigma$) of species $X$ relative to all species in icy/ocean and giant planets for all models, taking into account all chemical changes (CO:CO$_2$ variation and clathrate formation).}
\label{figp7}
\end{center}
\end{figure}

\paragraph{Composition of ices (molar fraction of species relative to all ices)} The chemical composition of species in planets is described in molar ratio relative to all volatile species (sum of H$_2$O, CO, CO$_2$, CH$_3$OH, CH$_4$, H$_2$S, N$_2$, and NH$_3$ molecules).
The presence of refractory organics does not change the relative (to H$_2$O) abundances of species $X$ in planets. So, the following section does not take into account this parameter for the study of the composition of volatile species. As discussed in paper 2, only the abundance of CO in planetesimals (CO:CO$_2$ ratio), the formation of clathrates, and the radiation (irradiated or non-irradiated models) change the chemical composition of planetesimals, and hence of planets. 
Figures \ref{figp8} and \ref{figp9} present, respectively, the average molar ratio of species H$_2$O, CH$_3$OH, NH$_3$, CO$_2$, H$_2$S, CH$_4$, CO, and N$_2$ in planets as a function of final semi-major axis a$_{pl}$ and the mass of the solid component M$^s_{pl}$, for irradiated (white area, right panel) or non-irradiated (dark area, left panel) models, taking into account all the chemical changes (CO:CO$_2$ variation, clathrate formation).
Figure \ref{figp7} compiles all the results by showing the average molar ratio ($\pm$2$\sigma$) of species $X$ relative to all volatile species in icy/ocean and giant planets, and for all models taking into account the chemical variation (formation of clathrates, CO:CO$_2$ variation).\\
Rocky planets (in area $\beta$) contain only a few ices that are mainly composed of H$_2$O molecules (see Fig.~\ref{figp8}), both for irradiated and non-irradiated models. So, the chemical composition of these planets is not discussed further. Planets that contain a mix of different ices are only found in the areas $\alpha$ and $\epsilon$. In these areas, the average molar ratio of H$_2$O (the most abundant volatile molecule) relative to all ices is equal at 63$\pm$14\% in icy/ocean planets and increase by about 5\% in giant planets of non-irradiated models (see Fig.~\ref{figp7}, area $\epsilon$). The standard deviation 2$\sigma$ takes into account the values of all non-irradiated models. Planets in irradiated models contain larger abundances (relative to all species) of H$_2$O with average values of 80$\pm$14 \% for icy/ocean planets, and 89$\pm$12\% for giant planets. This means that abundances of species more volatile than H$_2$O are depleted (relative to H$_2$O) in planets of irradiated models. As explained in paper 2, the irradiated models always contain larger abundances of H$_2$O relative to other species, but with continuous radial variations due to irradiation. This is due to a higher temperature in the disk which decreases the amount of highly volatile species incorporated in planetesimals as shown by figures 3 and 7 in paper 2.
Note, however, that the total amount of H$_2$O (and other volatile species) incorporated in irradiated planets decreases compared to the one in non-irradiated models. Remember that the amount of ices incorporated in planets decreases in irradiated compared to non-irradiated models (see figures \ref{figp4} and \ref{figp5}, and figures 8 and 11 for planetesimals in paper 2).
We also note that the relative (to all volatile species) abundance of H$_2$O in irradiated models shows radial variations due to the slow decrease of the temperature of the disk and the corresponding slow increase in relative (to all volatile species) abundances of all the species with the distance to the star (see figures 3, 7, and 8 in paper 2).

The molar fractions of the volatile species CO$_2$, CH$_3$OH, NH$_3$, and H$_2$S show relative variations of about 10\% (relative to their respecting average abundance) in icy/ocean planets of non-irradiated models (see Figures~\ref{figp8} and \ref{figp9}).
Their average abundance ($\pm$2$\sigma$) are roughly equal to 8-9$\pm$2\% for both CO$_2$ and CH$_3$OH, and 3.2$\pm$1\% and $\approx$1$\pm$0.2\%, respectively, for NH$_3$ and H$_2$S.  
In irradiated models, the average abundance of these species is equal to or lower by a few percentage points compared to non-irradiated models, but with higher standard deviation. The average abundance of the species CO$_2$, CH$_3$OH, NH$_3$, and H$_2$S in icy/ocean planets are roughly equal to 5.5$\pm$7\% for CO$_2$, 10$\pm$3\% for CH$_3$OH, and 2.6$\pm$2\% and $\approx$0.5$\pm$0.7\% for, respectively, NH$_3$ and H$_2$S.
In the giant planets, the abundances of the less volatile species shows larger depletion. 

In all models, giant planets are depleted in highly volatile species such as CH$_4$, CO, and N$_2$ compared to icy/ocean planets located far from the star ($\geq$10 AU). 
While the abundance of these species decrease by a few percentage points in giant planets of non-irradiated models, giant planets in irradiated disks are fully depleted (relative abundances inferior to 1\%) in high volatile species such as CH$_4$, CO, and N$_2$ (see Figures~\ref{figp9} and \ref{figp7}).
In icy/ocean planets (non-irradiated models), the average abundance of volatile molecules is equal to 1.7$\pm$1\% and 1.2$\pm$1\% for CH$_4$ and N$_2$, respectively. 
The CO abundance in planets (see Fig.~\ref{figp9}) vary with the CO abundance assumed in the solar nebula before the condensation/trapping of species (see paper 2). 
Taking into account all non-irradiated models, its average value is 13$\pm$16\% in icy/ocean planets. In giant planets, CO decreases to about 9$\pm$13\%.
In irradiated models (see Fig.~\ref{figp9}), planets present a lower abundance of highly volatile species with a higher standard deviation (see Fig.~\ref{figp7}).

\begin{figure*}
\begin{center}
\includegraphics[width=15.cm]{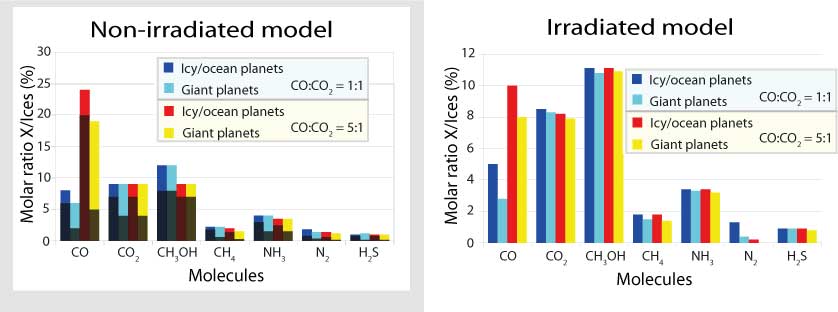}
\caption{Minimum (dark area) and maximum (light area) molar ratios of species $X$ relative to ices (the sum of H$_2$O, CO, CO$_2$, CH$_3$OH, CH$_4$, NH$_3$, N$_2$, and H$_2$S) in icy/ocean planets and giant planets for all models.}
\label{figp10}
\end{center}
\end{figure*}

Figure \ref{figp10} synthesizes all the results: it presents the minimum and maximum molar ratio of species $X$ (except H$_2$O) relative to all ices in planets for all models. 
 Table \ref{paramabundancesplanetsmassratio} summarizes the average molar fractions of species (relative to all species) for all models. \\

Icy/ocean planets have approximately the same (homogeneous) chemical composition as planetesimals (see paper 2), because they grow approximately at the same position as the formation of embryos (see Fig.~\ref{figp6}), while giant planets have moved in the protoplanetary disk and accreted planetesimals everywhere in the disk.
Giant planets have molecular abundances that can vary widely depending on the distance of formation of planets (inside or outside the water ice line) and their migration path in the protoplanetary disk.
If these planets are mainly composed of H$_2$O, CO$_2$, CH$_3$OH, and CO, they show high variations of all abundance species within this region, depending of the pathway of planets in the disk. 
Compared to the icy/ocean planets, a lot of giant planets are depleted in highly volatile molecules such as CH$_4$, N$_2$, and CO (see Figures~\ref{figp9} and \ref{figp7}, and Table~\ref{paramabundancesplanetsmassratio} for synthesis). From icy/ocean planets to giant planets, the abundance of CO and CH$_4$ is reduced by approximately 20-30\% in non-irradiated models. The H$_2$O abundance increases by about 5 wt\%. 
However, some of the giant planets are characterized by abundances of species similar to those of icy/ocean planets, such as NH$_3$, CH$_3$OH, and CO$_2$.

\begin{table*}
\centering 
\caption{Average molar ratio of species $X$ (with standard deviation 2$\sigma$) relative to all species in planets.}
\begin{tabular}{l|cc||cc}
\hline
\hline                              
Models	   & 	\multicolumn {2} {c||} {Non-irradiated}   				&	   \multicolumn {2} {c} {Irradiated}   \\
\hline  		
Planets	&  Icy/ocean  &  Giant													 & Icy/ocean   &  Giant \\
\hline	
X$_{pl}$/Ices$^*$ (\%)			 &   \multicolumn {2} {c||}{} &   \multicolumn {2} {c}{} \\
					
H$_2$O	 									 &	63 $\pm$ 14 & 68 $\pm$ 12 													& 80 $\pm$ 14 & 88.6 $\pm$ 12	\\ 
CO	                       &	13 $\pm$ 16 & 9 $\pm$ 13  												& 1.7 $\pm$ 6 & 0.07 $\pm$ 1 \\
CO$_2$	  								 &	8.5 $\pm$ 2 & 8.2 $\pm$ 2  												& 5.5 $\pm$ 7 & 1.7 $\pm$4.4	   \\
CH$_4$	                   &	1.7	$\pm$ 1  & 1.4 $\pm$ 1.2 					& 0.37  $\pm$ 1.2  & 0.03 $\pm$	 0.3 	 \\
H$_2$S	                   &	\multicolumn {2} {c||} {0.8 $\pm$ 0.2}  					& 0.48 $\pm$ 0.7 &  0.14 $\pm$ 0.4 	\\
N$_2$	                     &	 1.2 $\pm$ 1 & 0.6 $\pm$ 0.8 										&	0.1 $\pm$ 0.4  &  0	 \\
NH$_3$	                   &  \multicolumn {2} {c||} {3.2 $\pm$ 0.7} 						&	 	2.6 $\pm$ 2.3 & 	1.17 	$\pm$ 2.3 \\
CH$_3$OH                   &	8.4 $\pm$ 2  & 9 $\pm$ 1.4   											&  9.7 $\pm$ 3 & 8.3 $\pm$ 6.2	 \\
\hline
\hline
\end{tabular}
\begin{flushleft}
$^*$Ices= H$_2$O + CO + CO$_2$ + CH$_4$ +  H$_2$S +  N$_2$ +  NH$_3$ +  CH$_3$OH
\end{flushleft}
\label{paramabundancesplanetsmassratio}
\end{table*}

\paragraph{Abundances of the O, C, N , and S atoms relative to H$_2$}

\begin{figure}
\begin{center}
\includegraphics[width=8.cm]{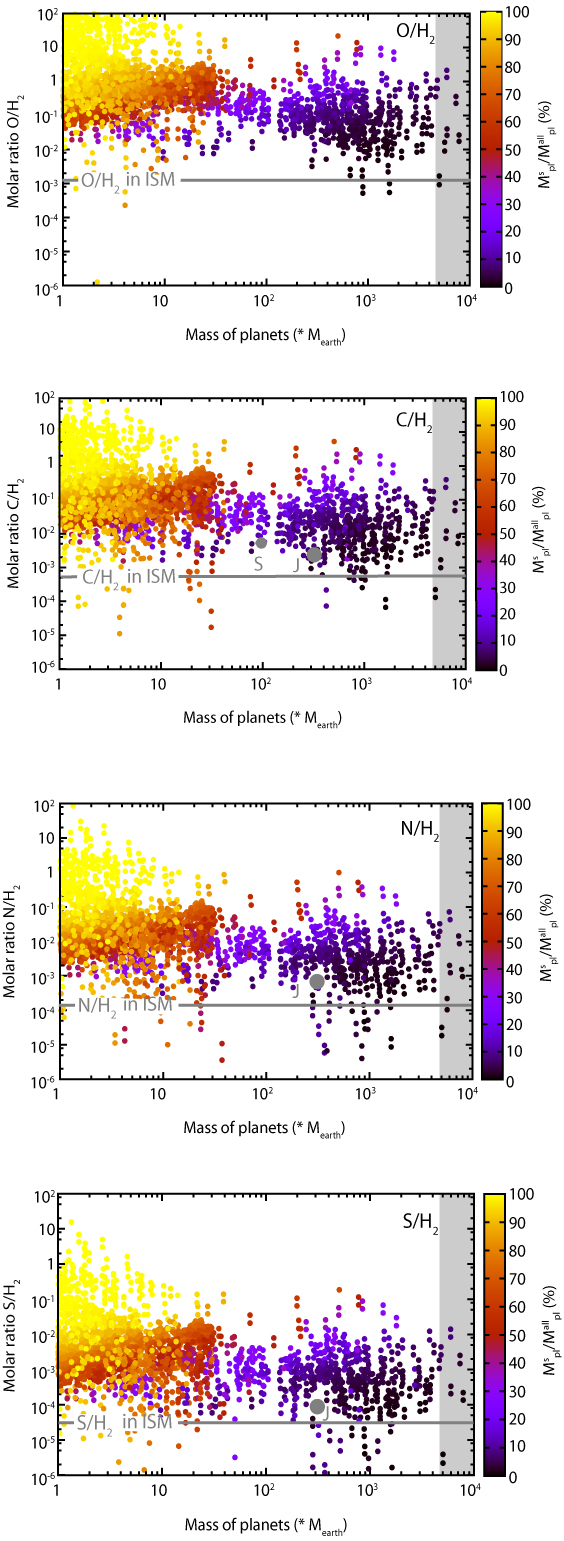}
\caption{Molar ratio of atoms O, C, N, and S relative to the total mass of gas H$_2$ in planets as a function of the total mass M$^{all}_{pl}$ of planets and mass ratio of solids M$^s_{pl}$ relative to total planetary mass M$^{all}_{pl}$, for all models. Horizontal gray lines represent abundances of atoms adopted in the ISM. Gray dots $J$ and $S$ represent, respectively, the abundances for Jupiter and Saturn. Data for Jupiter and Saturn provided by Wong et al. (2004), Flasar et al. (2005), Orton et al. (2005), and Briggs \& Sackett (1989). The light gray area on the right corresponds to the zone where the mass of planets is higher than 13 Jovian masses and where there could be more brown dwarfs than giant planets.}
\label{figp11}
\end{center}
\end{figure}

In order to compare observational chemical compositions of giant planets and exoplanets to our results, we provide the abundance of atoms O, C, N, and S (initially included in volatile species H$_2$O, CO, CO$_2$, CH$_3$OH, CH$_4$, NH$_3$, N$_2$, and H$_2$S) relative to H$_2$ in the atmospheres of planets (see Fig.~\ref{figp11}), assuming that all volatile species are in the gaseous state of the atmosphere of planets, whatever the position and surface temperatures of planets in the disk. We assume that ices initially incorporated in planetesimals are delivered as ices onto planets, but sublimate fully at the surface of planets after or during their accretion. 
Figure \ref{figp11} presents the molar ratio of atoms O, C, N, and S, relative to the total amount of H$_2$ in planets as a function of the total planetary mass M$^{all}_{pl}$ and the M$^s_{pl}$:M$^{all}_{pl}$ mass ratio in planets, for all models. Gray dots labeled $J$ and $S$ represent the abundances of atoms C, N, and S for planets Jupiter and Saturn, respectively (Wong et al. 2004; Fletcher et al. 2009).
For all atoms, the X:H$_2$ molar ratio in the atmosphere of planets is almost always greater than that in the molecular cloud that gave birth to the planetary system (ISM value), except for some planets where it can become inferior. 
For giant planets, the X:H$_2$ molar ratio can vary by several orders of magnitude as a function of the total mass of planets, i.e., the mass of gas (H$_2$ + He), the radiation, composition of planetesimals, and the assumption on the formation of clathrates and refractory organics in the disks. As can be seen in the figures, the fraction of gas (H$_2$ and He) increases (the mass ratio of solids M$^s_{pl}$ relative to the mass of planets M$^{all}_{pl}$ decreases) with the total mass of planets, a result typical of planets formed in the core-accretion model. The ratio decreases slightly with the increase of the total mass of planets, leading to ISM values of abundances of species for high masses of planets. 
 
 Our results are in good agreement with abundances of observed species C, N, and S for Jupiter and Saturn, even if our data are mostly above the observational data of the two planets. Our data were obtained assuming that all volatile species are in the gas phase and homogeneously distributed in the atmosphere of planets, leading to overestimation of the abundance that could be observed in Jovian planets and future extrasolar planets. The abundance of atoms observed in the atmospheres of Jovian planets could not represent the overall abundance in the planet: some chemical species could exist partially as solid state deeper in the interior of planets, and partially as gas state in the atmosphere of planets as a function of the temperature, pressure, and type of volatile molecules. Moreover, global circulations in the atmosphere of planets can lead to horizontal/vertical mixing and cause departures from equilibrium chemistry (Baraffe et al. 2010), which could lead to discrepancies between abundances of species in cool and hot regions of planets. 
Our model also allows us to reproduce the enrichment of noble gas Xe, Kr, and Ar in Jupiter.

\paragraph{Abundances of atoms O and C, and C:O ratio in volatile species} 

\begin{figure*}
\begin{center}
\includegraphics[width=15.cm, angle=0]{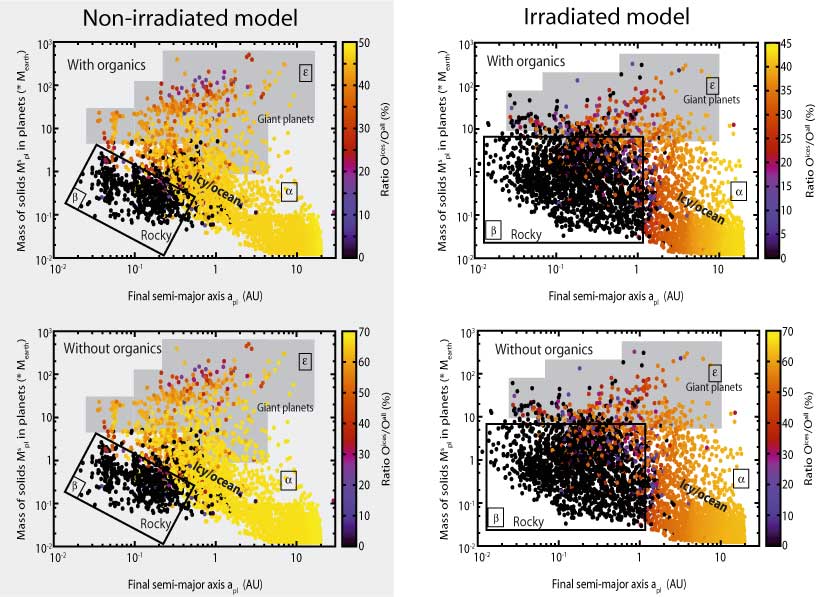}
\caption{Average ratio of atoms O$^{ices}$ in volatile molecules of planets relative to atoms O$^{all}$ in all solid components as a function of the solid mass M$^s_{pl}$ of planets and final semi-major axis a$_{pl}$, for non-irradiated (left panel) and irradiated (right panel) models, both with (upper panel) and without (lower panel) refractory organics, taking into account all chemical variations (CO:CO$_2$ variation and clathrate formation).}
\label{figp12}
\end{center}
\end{figure*}

\begin{figure}
\begin{center}
\includegraphics[width=8.cm, angle=0]{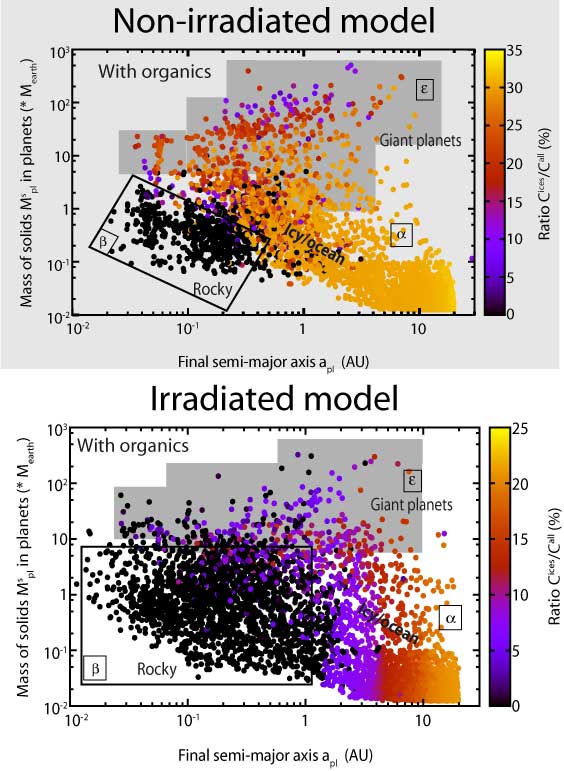}
\caption{Same as in Fig.~\ref{figp12} but for C atoms and with refractory organics only.}
\label{figp13}
\end{center}
\end{figure}

Figures \ref{figp12} and \ref{figp13} present the average ratio of the O (called O$^{ices}$) and C (called C$^{ices}$) atoms in volatile species relative to all atoms O (called O$^{all}$) and C (called C$^{all}$) respectively in planets, for non-irradiated and irradiated models, and with and without (only in Fig.\ref{figp12} for atoms O) refractory organics, taking into account all chemical variations (CO:CO$_2$ variation, and clathrate formation). 
For non-irradiated models, volatile species count for at least 46.3$\pm$10\% (resp. 66$\pm$13\%) of total atoms O, including minerals, organics, and volatile molecules in icy/ocean planets with (resp. without) refractory organics. They also include at least 29$\pm$10\% (resp. 100$\pm$6\%) of total atoms C for models with (resp. without) refractory organics. For giant planets, the average abundance of atoms is slighly lower by about 5-7\% with 41.5$\pm$14\% (resp. 60$\pm$18\%) for oxygen, and 22.7$\pm$12\% (resp. 100$\pm$12\%) for carbon in models with (resp. without) refractory organics. 
Planets without refractory organics show higher abundances of O and C in ices. This means that observations of volatile species in planets poor in refractory organics show abundances of atoms in good agreement with that of the full planet. 
For irradiated models, the abundances of atoms O and C show lower values with radial variations (see Table~\ref{paramabundancesatomsplanets} and Fig.~\ref{figp12}).

\begin{figure}
\begin{center}
\includegraphics[width=9.cm, angle=0]{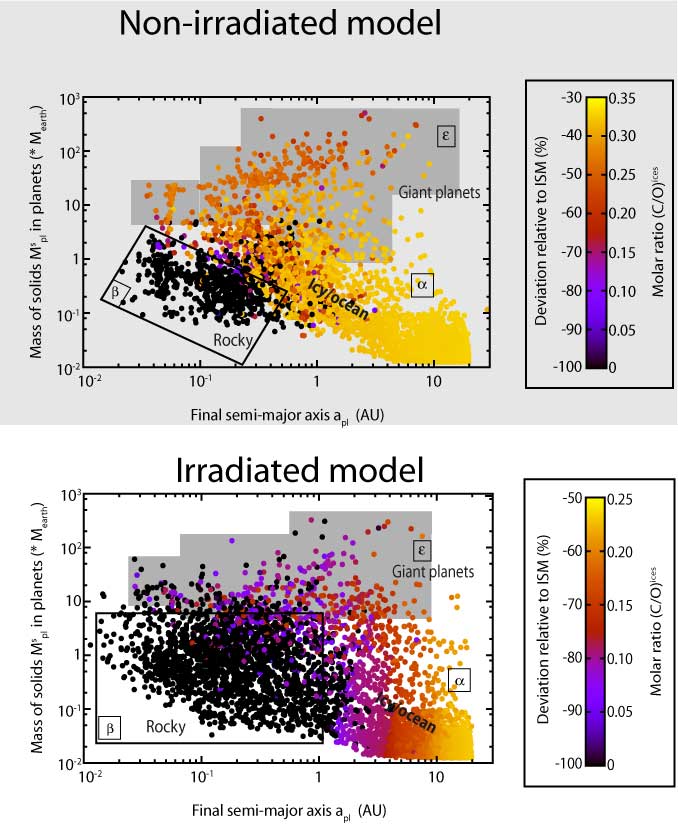}
\caption{Average molar ratio C:O in volatile species as a function of the solid mass M$^s_{pl}$ of planets and final semi-major axis a$_{pl}$. Non-irradiated (upper panel) and irradiated (lower panel) models taking into account all chemical variations (CO:CO$_2$ variation and clathrate formation).}
\label{figp14}
\end{center}
\end{figure}

One of the most important atomic ratios for the chemistry and mineralogy (see paper 1) of planets is the C/O ratio since it controls the distribution of carbon species. To characterize the depletion or enrichment of the C:O ratio in planets compared to the star, in Fig.~\ref{figp14} we plot the C:O ratio in the atmosphere (the main observable that we have on these objects) of planets (considering only ices, and assuming that all ices are in the gas phase) and the total C:O ratio of planets (considering ices, minerals, and refractory organics).
Figure \ref{figp14} presents the average C:O molar ratio in volatile molecules (right scale) and its deviation relative to the ISM value\footnote{The C:O molar ratio in the early stellar system (solar system composition) is 0.5 (Lodders, 2003). \label{ref_footnote}} (molar ratio $\frac{(C/O)^{ices} - 0.5}{0.5}$, left scale) as a function of the final semi-major axis a$_{pl}$ and the mass of solids M$^s_{pl}$ in planets, for non-irradiated (upper panel) and irradiated (lower panel) models, taking into account all chemical variations (CO:CO$_2$ variation and clathrate formation).
The C:O molar ratio in volatile molecules reaches 0.31$\pm$0.14 for icy/ocean planets with a deviation of -40$\mp$30\% relative to the ISM value$^{\ref{ref_footnote}}$ in non-irradiated models. For giant planets, the C:O ratio is slightly lower (0.27$\pm$0.12) with a deviation of -40$\mp$20\%. In irradiated models, all planets present lower values of the C:O molar ratio (values divided by about 2 compared to non-irradiated models), and with radial variations from 0 near the star to 0.17$\pm$0.11 (deviation of -60$\mp$10\%) far.
This means that the C:O ratio in volatile species that could be observed in the atmosphere of planets (and assuming that all volatile species are in the gas phase) range from deviations of -40$\pm$30\% to 0\% (the stellar value$^{\ref{ref_footnote}}$).

\begin{figure*}
\begin{center}
\includegraphics[width=15.cm, angle=0]{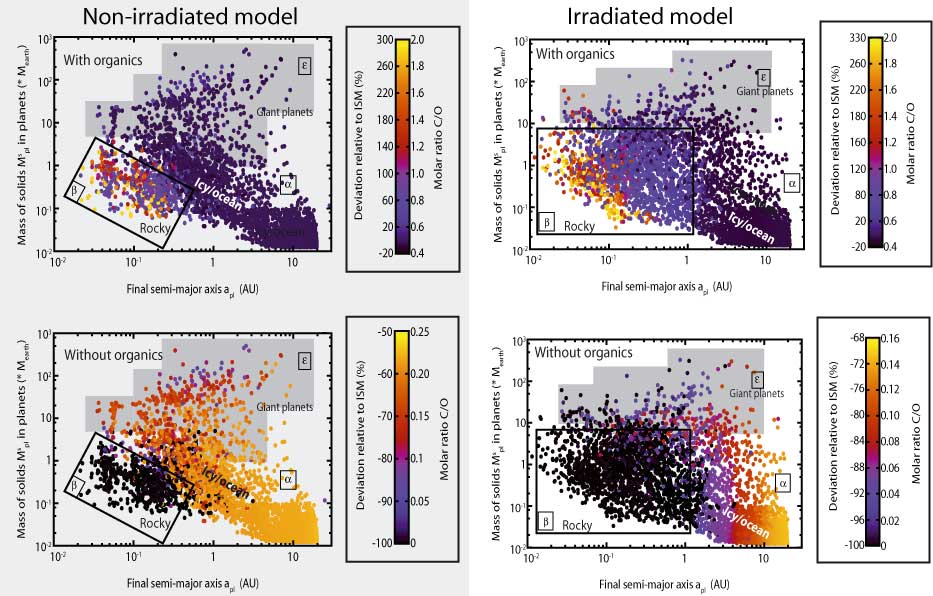}
\caption{Same as Fig.~\ref{figp14} but for the average molar ratio C:O in all solid components of planets.}
\label{figp15}
\end{center}
\end{figure*}

Figure \ref{figp15} presents the average C:O molar ratio in all elements (ices, minerals, refractory organics) of planets (right scale) and its deviation (molar ratio $\frac{(C/O)^{all} - 0.5}{0.5}$, left scale) relative to the ISM value$^{\ref{ref_footnote}}$ as a function of the final semi-major axis a$_{pl}$ and the mass of solids M$^s_{pl}$ in planets, for non-irradiated (left panel) and irradiated (right panel) models, with (upper panel) and without (lower panel) refractory organics, and taking into account all chemical variations (CO:CO$_2$ variation and clathrate formation). 
Icy/ocean and giant planets with refractory organics show C:O molar ratio roughly equal to 0.5$\pm$0.1 with a deviation of $\pm$10\% relative to the stellar value. Planets without refractory organics in non-irradiated models present lower C:O values at 0.21$\pm$0.09 (deviation of -60$\mp$20\%) for icy/ocean planets and 0.16$\pm$0.085 (deviation of -70$\mp$20\%) for giants. Planets in irradiated disks present values of the C/O that are approximately 2 and 3 times lower for icy/ocean and giant planets, respectively. If planets contain a large total amount of refractory organics, the C:O ratio should be of the same order of magnitude as the stellar value. Otherwise, the value should decrease by several tens of percent compared to the stellar value. We note that several rocky planets with refractory organics show high values of the C:O ratio, up to 3. This value corresponds to the C:O ratio in refractory organics. Without these refractory organics, the ratio decreases to 0 in rocky planets.
Table~\ref{paramabundancesatomsplanets} summarizes the average abundances (with standard deviation) of atoms O, C, and molar ratio C/O (in ices and all components) for all models, taking into account all chemical variations (formation of clathrates and CO:CO$_2$ variation).

\begin{table*}
\centering 
\caption{Average molar abundance (with standard deviation) of atoms $X$ in planets.}
\begin{tabular}{l|cc|cc||cc|cc}
\hline
\hline
Models	   & 	\multicolumn {4} {c||} {Non-irradiated}   				&	   \multicolumn {4} {c} {Irradiated}   \\
\hline
Planets	&  \multicolumn {2} {c|} {Icy/ocean}  &  \multicolumn {2} {c||} {Giant} 		&  \multicolumn {2} {c|} {Icy/ocean}  &  \multicolumn {2} {c} {Giant} \\
\hline			
Ref. organics                  	&	with 						 & without   			 &   with     & without     &	with 						 & without   			 &   with & without \\						
\hline
O$^{ices}$/O$^{all}$$^*$	(\%)	&	46.3$\pm$10 		 &	66$\pm$13		 &  41.5$\pm$14	& 60$\pm$18    &  38.5$\pm$12.7    &   59$\pm$14   & 29$\pm$20   &  45$\pm$30   \\ 
C$^{ices}$/C$^{all}$$^*$  (\%)	&	29 $\pm$10		 	 &	100$\pm$6   		     &	22.7$\pm$12	  &   100$\pm$12  		  &  14.4$\pm$11.6        &    98$\pm$26       &  7$\pm$9   &     95$\pm$42      \\
\hline
C/O$^{**}$ (ices)     &   \multicolumn {2} {c|} {0.31$\pm$0.14}    & \multicolumn {2} {c||} {0.27$\pm$0.12}   &   \multicolumn {2} {c|} {0.17$\pm$0.11}              &   \multicolumn {2} {c} {0.10$\pm$0.09}     \\
C/O$^{**}$	(all)  & 0.49$\pm$0.08 & 0.21$\pm$0.09   &  $\approx$0.50$\pm$0.11  &  0.16$\pm$0.085  & 0.47$\pm$0.054     &   0.10$\pm$0.08   & 0.50$\pm$0.12        &   0.051$\pm$0.06       \\
\hline
\end{tabular}
\begin{flushleft}
$^*$Average molar ratio of atoms X in ices relative to total atoms $X$ in planets (X$^{ices}_p$/X$^{all}_p$)\\
$^{**}$The molar stellar value of C/O in our study is 0.5.
\end{flushleft}
\label{paramabundancesatomsplanets}
\end{table*}

\section{Types of planets formed and the boundary for the models used \label{discussion}}
We now discuss the type of planets obtained, their possible physical evolution, and assumptions made in this study.

\subsection{Earth-like to super-Earth planets} 
	Planets with mass of solid components M$^{s}_{pl}$ varying from 0.01 M$_\oplus$ (dwarf planets) to about 10 M$_\oplus$ (usually designated super-Earth) are not massive enough to accrete a large gaseous envelope onto the central core (H$_2$ + He). These planet do not have a primitive gaseous atmosphere (H$_2$ + He) or the gaseous envelope of H$_2$ + He accreted by these planets corresponds to less than 10\% of the mass of solid components M$^{s}_{pl}$.
	\paragraph{Earth-like planets} 
	The rocky planets formed near stars ($a_{pl} \leq$ 0.4-2 AU) are mainly dry objects. However, these objects are not completely water-free.
For the Earth, H$_2$O represents  about 0.02\% of the core (Bell 2010). Despite its small relative abundance, water has dramatic effects on our planet's geology, geochemistry, climate, and biology and has important implications for planetary evolution and appearance of life (Bell 2010).
	  Most rocky planets formed in the simulation have final semi-major axes below 0.4-2 AU, but there are some rocky planets beyond 2 AU. After migration, some rocky planets are within the habitable zone, typically in the range 95-135\% of the midline, based mostly on analogy with our own solar system (Bell 2010).  
Post-formation degassing, comet accretion, or oxidization processes can only produce a tenuous gaseous atmosphere, with no significant consequences for the planet's contraction (Baraffe et al. 2010). 
	 
	 \paragraph{Icy/ocean planets}	 
	 Icy/ocean planets with mass of solid components M$^{s}_{pl}$ below about 5-10 M$_\oplus$ are located beyond 0.04-2 AU.
	   Some icy/ocean planets are within the inner and outer boundaries part of the habitable zone. These planets are composed of at least 23$\pm$10 wt\% of volatile molecules such as H$_2$O, CO$_2$, CO, CH$_3$OH, and NH$_3$, and have an ice/rock mass ratio varying from about 0.3$\pm$0.16 to 1.01$\pm$0.33.	Such values match those of some icy satelites and KBOs in the solar system: densities of Jovian and Saturn icy moons Ganymede, Callisto, and Titan, and Kuiper Belt objects Triton, Eris, and the Pluto-Charon binary ($\approx$ 1.8-2.3 g.cm$^{-3}$) suggest that rock and ices have comparable masses (Showman \& Malhotra 1999; Buie et al. 2006; Person et al. 2006; Brown \& Schaller 2007; McKinnon et al. 2008; Lunine et al. 2010; Sotin et al. 2010). 
	 Far away from the star where the temperature is low, planets are mainly icy with probably a tenuous atmosphere of highly volatile molecules such as CO and/or N$_2$ such as dwarf planets Charon and Pluto in the solar system or liquid or frozen planets such as Europa (one of Jupiter's moons) where tidal heating from Jupiter provides the energy source that would prevent a subsurface ocean from freezing solid (Bell 2010). \\
	 Near the star, where the temperature can be higher following the luminosity of the star, only the post-formation sublimation or liquefaction of ices due to temperatures of planetary surfaces can produce a gaseous atmosphere whose composition should be a function of the temperature of sublimation of volatile molecules and their escaping velocity in the planetary atmosphere. Such ocean planets present a large variation of abundances of volatile species probably due to the different origins of their formation and the pathways taken in the disk during their migration: H$_2$O can be the major volatile component or planets can include all volatile species in the same proportions as icy planets located far away from the star. We also note that warm temperatures could lead to chemical changes modifying the composition of the atmosphere. Such close-in and volatile rich super-Earths have already been observed such as GJ 1214b (6.55$\pm$0.98 M$_\oplus$, Miller-Ricci Kempton et al. 2012) and HD 97658b (8.2$\pm$1.2 M$_\oplus$, Dragomir et al. 2013). Note, however, that these planets orbit stars with lower masses and lower luminosities. Rogers \& Seager (2010) found that GJ 1214b, a planet orbiting a star of mass equal to 0.157M$_\odot$ (Charbonneau et al. 2009) and a luminosity equal to 0.3\% the Sun's, must be a water-rich planet composed of at least 60\% water by mass with a thick steam atmosphere to reproduce the planet's observed radius (2.68$\pm$0.13 R$_\oplus$, Charbonneau et al. 2009). However, an atmosphere composed of a combination of water vapor and hydrogen gas is also consistent with the planet's observed mass and radius, and these would require a smaller fraction of water to reproduce the planet's bulk density (Kempton et al. 2012). By examining all possible mixtures of water and H/He, Valencia et al. (2013) found that the bulk amount of H/He in GJ 1214b is at most 7\% by mass. Interestingly, our models, applied to sun mass and sun luminosity, show that some planets below 0.1 AU can have water mass ratio ranging from 10\% to 50\%. Applying our models to stars of lower mass and lower luminosity with different abundances of atoms could match the composition and position of planets observed more accurately.

 \subsection{Super-Earth to giant planets}
	 	Planets with mass of solid components M$^{s}_{pl}$ higher than 5-10 M$_\oplus$ (usually denominated as super-Earth) which are massive enough to accrete a large gaseous envelope onto the central core (H$_2$ + He) are labeled Neptune-like or gas giant planets. The gaseous envelope of H$_2$ + He accreted by these planets correspond to values much higher than 10\% of the mass of solid components M$^{s}_{pl}$.
	 	
	 \paragraph{Super-Earth planets to Neptune-like planets} 
	Planets with mass of solid components M$^{s}_{pl}$ varying from 5 M$_\oplus$ to about 20 M$_\oplus$ usually denominated as super-Earth down to Earth-like planets, are not massive enough to lead to rapid accretion (runaway regime) of a large gaseous envelope onto the central core. However, terrestrial planets can accrete a gaseous envelope of H$_2$ + He corresponding to up to 100 wt\% of the mass of solid components M$^{s}_{pl}$, leading to planets composed of 50 wt\% of solids (ices, minerals, refractory organics)  and 50 wt\% of gas H$_2$ + He. This leads to formation of Neptune- and Uranus-like planets. For these planets, the gaseous envelope essentially governs the gravothermal evolution of the planet (Bell 2010). However, differentiation, outgassing, and additional late heavy bombardment of icy planetesimals and subsequent accretion of chemical species could result in the enrichment of volatile molecules in their atmospheres over time, while H$_2$ could slowly be lost because of escape velocity and solar wind breakdown. This could then lead over time to the enrichment of the abundances of volatile molecules (relative to H$_2$) in the gaseous atmosphere of these planets. The abundance of species, i.e., the chemical reactions, in the gas envelope (dominated by H$_2$ and He) of these planets is temperature-pressure dependant and could change significantly with close-in planets ($\leq$ 0.1 AU) which could be affected by high temperatures.
	Close-in Neptune-like planets, produced in our model, and called as hot Neptunes, have already been observed, such as the planets GJ 436b (23 M$_\oplus$, Ag{\'u}ndez et al. 2014) and HAT-P-11b (23 M$_\oplus$, Lecavelier des Etangs et al. 2013). Note however that these planets orbit stars with significantly lower masses and luminosities, and different atom abundances. GJ 436b, a planet orbiting a star of mass equal to 0.41M$_\odot$ and a luminosity equal to 2.5\% the Sun's (Butler et al. 2004), at about 0.03 AU, has a planetary effective temperature of about 700-800 K (Deming et al. 2007; Demory et al. 2007; Stevenson et al. 2010) and an atmosphere rich in CO, but poor in CH$_4$ and H$_2$O (Stevenson et al. 2010; Madhusudhan \& Seager 2011; and Moses et al. 2013). Our model does not predict such a composition of planets, but chemical processes at high temperature\footnote{such as described by equations 1, 2, and 3 in the atmosphere of gas giant planets, see paper 2} that change the abundances of volatile molecules are not taken into account in this study since we produce only the original bulk composition derived from planetesimals.

	 \paragraph{Neptune-like to super-Jupiter planets} 
	Gas giant planets with mass of solids varying from 10-20 M$_\oplus$ to 500 M$_\oplus$ have a primitive gaseous atmosphere (H$_2$ + He), with mass ratio of solid components M$^{s}_{pl}$ relative to the total mass M$^{all}_{pl}$ of planets ranging from 60 wt\% to values lower than 10 wt\%. This leads to planets with ice/rock mass ratios varying from roughly 0.1 to 1.5.
	Our results for the ice/rock mass ratio in gas giant planets are slightly lower than those assumed in theoretical studies of giant planet formation as explained in paper 2. When gas water molecules condense, the solid surface density $\Sigma$ of the disk is assumed to increase by a factor of ~3-4 (Lecar et al. 2006; Encrenaz 2008). This assumption helps gas giant planets to form rapidly before the quick dissipation of the gas in the disk on a timescale of a few million years due to accretion onto the central protostar and evaporation from the protostar's irradiation. However, our results show that the ice/rock mass ratio in planetesimals (paper 2) and planets is on the order of magnitude 1$\pm$0.5, following the different assumptions on the thermodynamic conditions of formation of ices in the disks (see paper 2), leading to an increase in only the solid components by approximately 1.5-2.5 beyond the ice line, which is a factor lower by 2$\pm$0.5 than our best value compared to the one indicated by papers of planet formation (Hayashi 1981; Stevenson \& Lunine 1988; Encrenaz 2008). This discrepancy with our results can be explained mainly by the use by previous models (Hayashi 1981) of solar atomic abundances of Cameron (1970) that are higher by 50 - 90 \% for atoms C, O, and N compared to Lodders's (2003) abundances taken into account in this study. Moreover, Hayashi (1981) made some assumptions on the volatile species, taking into account only H$_2$O, CH$_4$, and NH$_3$, all condensing together at 170 K (see paper 2).
	
	 Theoretical models of Jupiter's interior predict that the total mass of heavy elements ranges from 20 M$_\oplus$ to 40 M$_\oplus$ (Saumon \& Guillot 2004; Nettelmann et al. 2008; Militzer et al. 2008, Helled \& Schubert 2009), leading to M$^{s}_{pl}$:M$^{all}_{pl}$ mass ratio ranging from 6\% to 13\% for this planet, in good agreement with results of the model of Alibert et al. (2013).
	However, our results on the ice/rock mass ratio are slightly different from those of the internal structure models of Uranus and Neptune (Podolak et al. 1995). Three-layer models with a central rocky core, an ice layer, and an outer H/He envelope suggest an overall composition of 25 wt\% of rocks, 60-70 wt\% of ices, and 5-15 wt\% of gaseous H/He (Podolak et al. 1991, Baraffe et al. 2010). Other solutions exist, as suggested by Podolak et al. (1995), assuming not a pure ice second layer, but a mixture of ices, rock (minerals + refractory organics), and gas. However, the ice/rock mass ratio remains at about 2.5 for both planets Uranus and Neptune in these models, $\approx$ 1.5 times higher that our best results. 
	 In the solar system, the largest reservoirs of water should be Jupiter and Saturn which represent 91\% of the planetary mass in our solar system (Fortney \& Nettelmann 2010), and Uranus and Neptune (7\% of the planetary mass in our solar system). However, there are no data available on the abundance of tropospheric or deeper ices (mainly H$_2$O) on these planets. The bulk compositions of these planets are not well constrained (Podolak et al. 1995; Marley et al. 1995; Helled et al. 2011; Podolak \& Helled 2012). In fact, it is still unclear what the mass fraction of ices is in Uranus and Neptune, despite their categorization as icy planets (Podolak \& Helled 2012). \\

	 Measurements by the Galileo probe mass spectrometer suggest that the atmosphere of Jupiter is enriched in heavy elements by a factor of up to 3 compared with the Sun (Young 2003, Helled \& Schubert 2009). Our results suggest that the abundances of atoms (relative to H$_2$) in giant planets are almost higher that the stellar values (up to 3 orders of magnitude for low mass planets), but decrease with an increase in the total mass of planets, leading to stellar values for large planet masses which means there could be more brown dwarfs than giant planets for total masses higher than 13 Jupiter masses. The abundances of species (relative to H$_2$) in the atmosphere of giant planets determined in this work are upper limits. First, we assumed that all volatile species incorporated as ices in planets were in the gaseous state of the atmosphere of planets although a part could be in the solid state, deeper in the planets, as a function of the temperature, pressure, and type of volatile molecules. Second, the heating of the atmosphere of planets (after migration toward the star) near the star could trigger slow chemical reactions such as described in Eqs.~1, 2, and 3 in paper 2, leading to the enrichment or depletion of species in the atmosphere of planets. Moreover, global circulations of the atmosphere of planets can lead to discrepancies of abundances of species between cool and hot regions of the atmosphere.

The model of planet formation used in this work (Alibert et al. 2013) assumes small planetesimals that keep the same radius (about 100 m). As shown in Fortier et al. (2013), the formation of giant planets is favored by the accretion of small planetesimals which do not evolve with time and leads to the formation of massive planets with masses higher than 500 M$_\oplus$. Increasing the radius of planetesimals reduces severely the accretion rate and the growth of planets (see Alibert et al. 2013; Fortier et al. 2013). Assuming a size distribution for planetesimals and a radius evolution of planetesimals (increase due to accretion of smaller objects, and decrease due to the sublimation of ices when shifted toward the star) could change the mass of planets formed and presented in this study. Such considerations will be the subject of future works.

It is important to remember that the chemical compositions of planets given in this study do not take into account loss of material due to collisional heating and breaking of bodies; size of bodies and temperature of sublimation of volatile molecules (evaporation process, escape velocity of volatile molecules); tidal and radioactive heating and geysers as observed in the satellites Enceladus and Europa; and chemical processes \footnote{such as described by equations 1, 2, and 3 in the atmosphere of gas giant planets, see paper 2} changing the abundances of volatile molecules. 
In addition, this study does not consider the radial migration of grains and small planetesimals ($\leq$ 1 km) in the disk due to the effect of gas drag and gravitational torques. 
We computed the chemical composition of ices in planetesimals and planets by using the thermodynamic conditions derived from an $\alpha$ disk both with and without irradiation, and for solar conditions (mass, luminosity, and composition of atoms).
The irradiated and non-irradiated models used in this study do not properly model the physics in the disks. However, they allow us to frame the thermodynamic conditions of ice formation in the cooling disk and migration of planets, respectively, near and far away from the star.
So, the abundance of volatile molecules of planets calculated in this work constitute an upper (resp. lower) limit for non-irradiated (resp. irradiated) models in solar conditions. 
The gas phase incorporated by giant planets and assumed to be composed only of H$_2$ and He, should be composed of other volatile species as suggested by Oberg et al. (2011). Calculating the gas composition of volatile molecules incorporated in giant planets could change slightly the chemical gas composition of these planets supposed to be derived directly from ice species in this study (see Fig.~\ref{figp11}) and change their carbon to oxygen ratio. We currently extend our work in this direction.
Finally, the chemical composition of planets and icy satellites are derived from average values of abundances of species observed in the ISM. Abundances of species can change up to several tens of percentage points for some molecules such as CO$_2$, CH$_3$OH, and NH$_3$ (see paper 2). A system with lower extremes or larger abundances of species compared to average values taken in this study could change the chemical composition of planets given in this paper.\\
In order to better constrain the chemical composition of planets, future work will focus on the nonsolar composition of atoms (C/O variation), different masses and luminosities of star, and will include the radial migration of grains and planetesimals in the disk. The computation of radii of planets in a self-consistent way using these results will also be done in future work.

\section{Summary \label{conclusion}}

In this work, we have computed the composition of volatile molecules incorporated in planets after migration in the disk, using 16 different physico-chemical conditions. To do this, we proceeded in two steps. In the first paper (see paper 2), we used models of disk accretion and ice formation to calculate the abundance of ices incorporated in planetesimals for different initial physico-chemical conditions, varying the abundance of CO in planetesimals, i.e., the disk, taking into account, or not, the presence of refractory organics, clathrates, star radiation, and mass of the disk. In this paper, starting from the chemical composition calculated in the disk for different surface densities, we used the model of planet formation and migration of Alibert et al. (2013) to determine a wide variety of chemical composition of planets listed in three groups:
\begin{itemize}
 \renewcommand{\labelitemi}{$\bullet$}
 \item Giant planets with mass of solids 5-10 M$_\oplus$ $\leq$ M$^s_{pl}$ $\leq$ 500 M$_\oplus$.
 \item Icy/ocean planets with mass of solids M$^s_{pl}$ $\leq$ 5-10 M$_\oplus$.
 \item Rocky planets with mass of solids M$^s_{pl}$ $\leq$ 5-10 M$_\oplus$, and mainly located inside 2 AU.
\end{itemize}
 
The main results regarding the chemical composition of these planets are described below.
 \begin{enumerate}
 \item Icy/ocean planets do not migrate much and retain approximately the same chemical composition as planetesimals beyond the ice lines.
 \item Giant planets show high variations of abundances of all volatile species and some of them are depleted in ices relative to icy/ocean planets. Many giant planets in non-irradiated models have lower abundances of highly volatile species such as CH$_4$, CO, and N$_2$, and most of them in irradiated models are depleted in these species compared to icy/ocean planets.
 \item Ice/rock and Ice/all elements ratio:
\begin{itemize}
\renewcommand{\labelitemi}{$\bullet$}
	\item The mass fraction of ices varies from 23$\pm$10 to 50$\pm$12wt\% in icy/ocean planets. This value decreases to less than 0.01 wt\% in rocky planets and can reach up to 43$\pm$17\% in giant planets.
 	\item The ice/rock mass ratio in planets varies from 0 (rocky planets) to 1.5 (icy/ocean and gas giant planets).
\end{itemize}
	\item Chemical composition of planets:
\begin{itemize}
\renewcommand{\labelitemi}{$\bullet$}
  \item The dominant chemical species in ices remains H$_2$O (at least 63$\pm$14\% in mol relative to all ices) whatever the type of planets formed and assumptions on the disk, the chemical composition, and evolution.
	\item The three dominant molecules after H$_2$O are carbon-bearing species such as CO, CO$_2$, and CH$_3$OH. 
	\item The dominant nitrogen-bearing species remains NH$_3$.
\end{itemize}
	\item Atomic composition of planets:
\begin{itemize}
\renewcommand{\labelitemi}{$\bullet$}
	\item The abundance of atoms O, C, N, and S (relative to H$_2$) in the atmosphere of gas giant planets is almost greater than the stellar values for most planets, but decreases slightly with the increase of the total mass of planets.
	\item Our results on the abundance of atoms C, N, and S (relative to H$_2$) are in good agreement with the abundances observed for Jupiter and Saturn.
	\item Taking into account all physico-chemical variations, the C:O molar ratio in volatile molecules reaches up to 0.31$\pm$0.14 for icy/ocean planets, and 0.27$\pm$0.12 for gas giant planets, about 40 - 50\% of the stellar value.
	\item The total C:O molar ratio in planets (considering ices, minerals, and refractory organics) reaches up to 0.5$\pm$0.1 considering refractory organics, and 0.21$\pm$0.09 without, about 0 - 60\% of the stellar value.
\end{itemize}
	\end{enumerate}

\begin{acknowledgements}
This work has been supported by the Swiss National Science Foundation, the Center for Space and Habitibility of the University of Bern and the European Research Council under grant 239605.
\end{acknowledgements}


\begin{thebibliography}{}

\bibitem[Ag{\'u}ndez et al.(2014)]{2014ApJ...781...68A} Ag{\'u}ndez, M., Venot, O., Selsis, F., \& Iro, N.\ 2014, \apj, 781, 68
%The Puzzling Chemical Composition of GJ 436b's Atmosphere: Influence of Tidal Heating on the Chemistry


\bibitem[Alibert et al.(2013)]{} Alibert, Y., Carron, F., Fortier, A., et al.\ 2013, \aap, submitted 
%Theoretical models of planetary system formation: mass vs semi-major axis


\bibitem[Alibert et al.(2005)]{2005A&A...434..343A} Alibert, Y., Mordasini, C., Benz, W., \& Winisdoerffer, C.\ 2005, \aap, 434, 343 
%Models of giant planet formation with migration and disc evolution

\bibitem[Baraffe et al.(2010)]{2010RPPh...73a6901B} Baraffe, I., Chabrier, G., \& Barman, T.\ 2010, Reports on Progress in Physics, 73, 016901 
%The physical properties of extra-solar planets

\bibitem[Barclay et al.(2013)]{2013Natur.494..452B} Barclay, T., Rowe, J.~F., Lissauer, J.~J., et al.\ 2013, \nat, 494, 452 
%A sub-Mercury-sized exoplanet

\bibitem[Barman(2008)]{2008ApJ...676L..61B} Barman, T.~S.\ 2008, \apjl, 676, L61 
%On the Presence of Water and Global Circulation in the Transiting Planet HD 189733b

\bibitem[Barman(2007)]{2007ApJ...661L.191B} Barman, T.\ 2007, \apjl, 661, L191 
%Identification of Absorption Features in an Extrasolar Planet Atmosphere

\bibitem[Beaulieu et al.(2010)]{2010MNRAS.409..963B} Beaulieu, J.~P., Kipping, D.~M., Batista, V., et al.\ 2010, \mnras, 409, 963 
%Water in the atmosphere of HD 209458b from 3.6-8 µm IRAC photometric observations in primary transit

\bibitem[Beaulieu et al.(2008)]{2008ApJ...677.1343B} Beaulieu, J.~P., Carey, S., Ribas, I., \& Tinetti, G.\ 2008, \apj, 677, 1343 
%Primary Transit of the Planet HD 189733b at 3.6 and 5.8 µm

\bibitem[Bell(2010)]{2010HiA....15...29B} Bell, J.~F.\ 2010, Highlights of Astronomy, 15, 29 
%Water on planets

\bibitem[Bockel{\'e}e-Morvan et al.(2004)]{2004come.book..391B} Bockel{\'e}e-Morvan, D., Crovisier, J., Mumma, M.~J., \& Weaver, H.~A.\ 2004, Comets II, 391
%The Composition of Cometary Volatiles

\bibitem[Bond et al.(2010)]{2010ApJ...715.1050B} Bond, J.~C., O'Brien, D.~P., \& Lauretta, D.~S.\ 2010, \apj, 715, 1050 
%The Compositional Diversity of Extrasolar Terrestrial Planets. I. In Situ Simulations

\bibitem[Brown \& Schaller(2007)]{2007Sci...316.1585B} Brown, M.~E., \& Schaller, E.~L.\ 2007, Science, 316, 1585
%The Mass of Dwarf Planet Eris 

\bibitem[Buie et al.(2006)]{2006AJ....132..290B} Buie, M.~W., Grundy, W.~M., Young, E.~F., Young, L.~A., \& Stern, S.~A.\ 2006, \aj, 132, 290
%Orbits and Photometry of Pluto's Satellites: Charon, S/2005 P1, and S/2005 P2
 
\bibitem[Burrows et al.(2008)]{2008ApJ...678.1436B} Burrows, A., Budaj, J., \& Hubeny, I.\ 2008, \apj, 678, 1436 
%Theoretical Spectra and Light Curves of Close-in Extrasolar Giant Planets and Comparison with Data
 
\bibitem[Burrows et al.(2007)]{2007ApJ...668L.171B} Burrows, A., Hubeny, I., Budaj, J., Knutson, H.~A., \& Charbonneau, D.\ 2007, \apjl, 668, L171 
%Theoretical Spectral Models of the Planet HD 209458b with a Thermal Inversion and Water Emission Bands 
 
\bibitem[Burrows et al.(2005)]{2005ApJ...625L.135B} Burrows, A., Hubeny, I., \& Sudarsky, D.\ 2005, \apjl, 625, L135 
%A Theoretical Interpretation of the Measurements of the Secondary Eclipses of TrES-1 and HD 209458b 
 
\bibitem[Butler et al.(2004)]{2004ApJ...617..580B} Butler, R.~P., Vogt, S.~S., Marcy, G.~W., et al.\ 2004, \apj, 617, 580
%A Neptune-Mass Planet Orbiting the Nearby M Dwarf GJ 436 

\bibitem[Charbonneau et al.(2009)]{2009Natur.462..891C} Charbonneau, D., Berta, Z.~K., Irwin, J., et al.\ 2009, \nat, 462, 891
%A super-Earth transiting a nearby low-mass star

\bibitem[Charbonneau et al.(2008)]{2008ApJ...686.1341C} Charbonneau, D., Knutson, H.~A., Barman, T., et al.\ 2008, \apj, 686, 1341 
%The Broadband Infrared Emission Spectrum of the Exoplanet HD 189733b

\bibitem[Charbonneau et al.(2005)]{2005ApJ...626..523C} Charbonneau, D., Allen, L.~E., Megeath, S.~T., et al.\ 2005, \apj, 626, 523 
%Detection of Thermal Emission from an Extrasolar Planet

\bibitem[Charbonneau et al.(2002)]{2002ApJ...568..377C} Charbonneau, D., Brown, T.~M., Noyes, R.~W., \& Gilliland, R.~L.\ 2002, \apj, 568, 377 
%Detection of an Extrasolar Planet Atmosphere
 
\bibitem[Deming et al.(2007)]{2007ApJ...667L.199D} Deming, D., Harrington, J., Laughlin, G., et al.\ 2007, \apjl, 667, L199 
%Spitzer Transit and Secondary Eclipse Photometry of GJ 436b

\bibitem[Demory et al.(2007)]{2007A&A...475.1125D} Demory, B.-O., Gillon, M., Barman, T., et al.\ 2007, \aap, 475, 1125
%Characterization of the hot Neptune GJ 436 b with Spitzer and ground-based observations

\bibitem[D{\'e}sert et al.(2009)]{2009ApJ...699..478D} D{\'e}sert, J.-M., Lecavelier des Etangs, A., H{\'e}brard, G., et al.\ 2009, \apj, 699, 478
%Search for Carbon Monoxide in the Atmosphere of the Transiting Exoplanet HD 189733b

\bibitem[Doty et al.(2002)]{2002A&A...389..446D} Doty, S.~D., van Dishoeck, E.~F., van der Tak, F.~F.~S., \& Boonman, A.~M.~S.\ 2002, \aap, 389, 446 
%Chemistry as a probe of the structures and evolution of massive star-forming regions

\bibitem[Dragomir et al.(2013)]{2013ApJ...772L...2D} Dragomir, D., Matthews, J.~M., Eastman, J.~D., et al.\ 2013, \apjl, 772, L2
%MOST Detects Transits of HD 97658b, a Warm, Likely Volatile-rich Super-Earth


\bibitem[Encrenaz(2008)]{2008ARA&A..46...57E} Encrenaz, T.\ 2008, \araa, 46, 57 
%Water in the Solar System

\bibitem[Fletcher et al.(2009)]{2009Icar..199..351F} Fletcher, L.~N., Orton, G.~S., Teanby, N.~A., Irwin, P.~G.~J., \& Bjoraker, G.~L.\ 2009, \icarus, 199, 351
%Methane and its isotopologues on Saturn from Cassini/CIRS observations

\bibitem[Fortier et al.(2013)]{2013A&A...549A..44F} Fortier, A., Alibert, Y., Carron, F., Benz, W., \& Dittkrist, K.-M.\ 2013, \aap, 549, A44
%Planet formation models: the interplay with the planetesimal disc

\bibitem[Fortney \& Nettelmann(2010)]{2010SSRv..152..423F} Fortney, J.~J., \& Nettelmann, N.\ 2010, \ssr, 152, 423
%The Interior Structure, Composition, and Evolution of Giant Planets

\bibitem[Fortney \& Marley(2007)]{2007ApJ...666L..45F} Fortney, J.~J., \& Marley, M.~S.\ 2007, \apjl, 666, L45
%Analysis of Spitzer Spectra of Irradiated Planets: Evidence for Water Vapor?

\bibitem[Gautier et al.(2001)]{2001ApJ...550L.227G} Gautier, D., Hersant, F., Mousis, O., \& Lunine, J.~I.\ 2001a, \apjl, 550, L227 
%Enrichments in Volatiles in Jupiter: A New Interpretation of the Galileo Measurements

\bibitem[Gautier et al.(2001)]{2001ApJ...559L.183G} Gautier, D., Hersant, F., Mousis, O., \& Lunine, J.~I.\ 2001b, \apjl, 559, L183 
%erratum: Enrichments in Volatiles in Jupiter: A New Interpretation of the Galileo Measurements

\bibitem[Griffith et al.(2011)]{2011epsc.conf..140G} Griffith, C.~A., Tinetti, G., Swain, M.~R., et al.\ 2011, EPSC-DPS Joint Meeting 2011, 140 
%Observational constraints on the composition of exoplanets

\bibitem[Grillmair et al.(2008)]{2008Natur.456..767G} Grillmair, C.~J., Burrows, A., Charbonneau, D., et al.\ 2008, \nat, 456, 767
%Strong water absorption in the dayside emission spectrum of the planet HD189733b

\bibitem[Hayashi(1981)]{1981PThPS..70...35H} Hayashi, C.\ 1981, Progress of Theoretical Physics Supplement, 70, 35 
%Structure of the Solar Nebula, Growth and Decay of Magnetic Fields and Effects of Magnetic and Turbulent Viscosities on the Nebula

\bibitem[Helled et al.(2011)]{2011ApJ...726...15H} Helled, R., Anderson, J.~D., Podolak, M., \& Schubert, G.\ 2011, \apj, 726, 15 
%Interior Models of Uranus and Neptune

\bibitem[Helled \& Schubert(2009)]{2009ApJ...697.1256H} Helled, R., \& Schubert, G.\ 2009, \apj, 697, 1256 
%Heavy-element Enrichment of a Jupiter-mass Protoplanet as a Function of Orbital Location

\bibitem[Hersant et al.(2004)]{2004P&SS...52..623H} Hersant, F., Gautier, D., \& Lunine, J.~I.\ 2004, \planss, 52, 623 
%Enrichment in volatiles in the giant planets of the Solar System

\bibitem[Johnson et al.(2012)]{2012ApJ...757..192J} Johnson, T.~V., Mousis, O., Lunine, J.~I., \& Madhusudhan, N.\ 2012, \apj, 757, 192 
%Planetesimal Compositions in Exoplanet Systems

\bibitem[Lecar et al.(2006)]{2006ApJ...640.1115L} Lecar, M., Podolak, M., Sasselov, D., \& Chiang, E.\ 2006, \apj, 640, 1115 
%On the Location of the Snow Line in a Protoplanetary disc

\bibitem[Lecavelier des Etangs et al.(2013)]{2013A&A...552A..65L} Lecavelier des Etangs, A., Sirothia, S.~K., Gopal-Krishna, \& Zarka, P.\ 2013, \aap, 552, A65 
%Hint of 150 MHz radio emission from the Neptune-mass extrasolar transiting planet HAT-P-11b

\bibitem[Lee et al.(2012)]{2012MNRAS.420..170L} Lee, J.-M., Fletcher, L.~N., \& Irwin, P.~G.~J.\ 2012, \mnras, 420, 170
%Optimal estimation retrievals of the atmospheric structure and composition of HD 189733b from secondary eclipse spectroscopy

\bibitem[Lodders(2003)]{2003ApJ...591.1220L} Lodders, K.\ 2003, \apj, 591, 1220 
%Solar System Abundances and Condensation Temperatures of the Elements

\bibitem[Lunine(2011)]{2011MmSAI..82..368L} Lunine, J.~I.\ 2011, \memsai, 82, 368 
%Extrasolar planets: the final frontier

\bibitem[Lunine et al.(2010)]{2010tfch.book...35L} Lunine, J., Choukroun, M., Stevenson, D., \& Tobie, G.\ 2010, Titan from Cassini-Huygens, 35
%The Origin and Evolution of Titan

\bibitem[Madhusudhan(2012)]{2012ApJ...758...36M} Madhusudhan, N.\ 2012, \apj, 758, 36 
%C/O Ratio as a Dimension for Characterizing Exoplanetary Atmospheres

\bibitem[Madhusudhan et al.(2011)]{2011Natur.469...64M} Madhusudhan, N., Harrington, J., Stevenson, K.~B., et al.\ 2011, \nat, 469, 64 
%A high C/O ratio and weak thermal inversion in the atmosphere of exoplanet WASP-12b

\bibitem[Madhusudhan \& Seager(2011)]{2011ApJ...729...41M} Madhusudhan, N., \& Seager, S.\ 2011, \apj, 729, 41
%High Metallicity and Non-equilibrium Chemistry in the Dayside Atmosphere of hot-Neptune GJ 436b

\bibitem[Madhusudhan \& Seager(2010)]{2010EGUGA..1214364M} Madhusudhan, N., \& Seager, S.\ 2010, EGU General Assembly Conference Abstracts, 12, 14364
%Temperature and abundance retrieval for exoplanet atmospheres

\bibitem[Madhusudhan \& Seager(2009)]{2009ApJ...707...24M} Madhusudhan, N., \& Seager, S.\ 2009, \apj, 707, 24
%A Temperature and Abundance Retrieval Method for Exoplanet Atmospheres


\bibitem[Marboeuf et al.(2014)]{} Marboeuf, U., Thiabaud, A., Alibert A., Cabral, N., \& Benz, W.\ 2014, \aap, submitted
%From setallar nebula to planetesimals

\bibitem[Marboeuf et al.(2008)]{2008ApJ...681.1624M} Marboeuf, U., Mousis, O., Ehrenreich, D., Alibert, Y., Cassan, A., Wakelam, V., \& Beaulieu, J.-P.\ 2008, \apj, 681, 1624 

\bibitem[Marley et al.(1995)]{1995JGR...10023349M} Marley, M.~S., G{\'o}mez, P., \& Podolak, M.\ 1995, \jgr, 100, 23349 
%Monte Carlo interior models for Uranus and Neptune

\bibitem[McKinnon et al.(2008)]{2008ssbn.book..213M} McKinnon, W.~B., Prialnik, D., Stern, S.~A., \& Coradini, A.\ 2008, The Solar System Beyond Neptune, 213
%Structure and Evolution of Kuiper Belt Objects and Dwarf Planets

\bibitem[Militzer et al.(2008)]{2008ApJ...688L..45M} Militzer, B., Hubbard, W.~B., Vorberger, J., Tamblyn, I., \& Bonev, S.~A.\ 2008, \apjl, 688, L45 
%A Massive Core in Jupiter Predicted from First-Principles Simulations

\bibitem[Miller-Ricci Kempton et al.(2012)]{2012ApJ...745....3M} Miller-Ricci Kempton, E., Zahnle, K., \& Fortney, J.~J.\ 2012, \apj, 745, 3 
%The Atmospheric Chemistry of GJ 1214b: Photochemistry and Clouds


\bibitem[Mordasini et al.(2012)]{2012A&A...547A.111M} Mordasini, C., Alibert, Y., Klahr, H., \& Henning, T.\ 2012a, \aap, 547, A111 
%Characterization of exoplanets from their formation. I. Models of combined planet formation and evolution

\bibitem[Mordasini et al.(2012)]{2012A&A...547A.112M} Mordasini, C., Alibert, Y., Georgy, C., et al.\ 2012b, \aap, 547, A112
%Characterization of exoplanets from their formation. II. The planetary mass-radius relationship

\bibitem[Mordasini et al.(2009)]{2009A&A...501.1139M} Mordasini, C., Alibert, Y., \& Benz, W.\ 2009, \aap, 501, 1139 
%Extrasolar planet population synthesis. I. Method, formation tracks, and mass-distance distribution

\bibitem[Moses et al.(2013)]{2013ApJ...763...25M} Moses, J.~I., Madhusudhan, N., Visscher, C., \& Freedman, R.~S.\ 2013, \apj, 763, 25
%Chemical Consequences of the C/O Ratio on Hot Jupiters: Examples from WASP-12b, CoRoT-2b, XO-1b, and HD 189733b

\bibitem[Mousis et al.(2010)]{2010FaDi..147..509M} Mousis, O., Lunine, J.~I., Picaud, S., \& Cordier, D.\ 2010, Faraday Discussions, 147, 509
%Volatile inventories in clathrate hydrates formed in the primordial nebula

\bibitem[Mumma \& Charnley(2011)]{2011ARA&A..49..471M} Mumma, M.~J., \& Charnley, S.~B.\ 2011, \araa, 49, 471 
%The Chemical Composition of Comets—Emerging Taxonomies and Natal Heritage

\bibitem[Nettelmann et al.(2008)]{2008ApJ...683.1217N} Nettelmann, N., Holst, B., Kietzmann, A., et al.\ 2008, \apj, 683, 1217 
%Ab Initio Equation of State Data for Hydrogen, Helium, and Water and the Internal Structure of Jupiter


\bibitem[Nomura \& Millar(2004)]{2004A&A...414..409N} Nomura, H., \& Millar, T.~J.\ 2004, \aap, 414, 409
%The physical and chemical structure of hot molecular cores


\bibitem[{\"O}berg et al.(2011)]{2011ApJ...743L..16O} {\"O}berg, K.~I., Murray-Clay, R., \& Bergin, E.~A.\ 2011, \apjl, 743, L16 
%The Effects of Snowlines on C/O in Planetary Atmospheres

\bibitem[Person et al.(2006)]{2006AJ....132.1575P} Person, M.~J., Elliot, J.~L., Gulbis, A.~A.~S., et al.\ 2006, \aj, 132, 1575
%Charon's Radius and Density from the Combined Data Sets of the 2005 July 11 Occultation

\bibitem[Podolak \& Helled(2012)]{2012ApJ...759L..32P} Podolak, M., \& Helled, R.\ 2012, \apjl, 759, L32
%What Do We Really Know about Uranus and Neptune?

\bibitem[Podolak et al.(1995)]{1995P&SS...43.1517P} Podolak, M., Weizman, A., \& Marley, M.\ 1995, \planss, 43, 1517 
%Comparative models of Uranus and Neptune

\bibitem[Podolak et al.(1991)]{1991uran.book...29P} Podolak, M., Hubbard, W.~B., \& Stevenson, D.~J.\ 1991, Uranus, 29 
%Model of Uranus' interior and magnetic field

\bibitem[Rogers \& Seager(2010)]{2010ApJ...716.1208R} Rogers, L.~A., \& Seager, S.\ 2010, \apj, 716, 1208 
%Three Possible Origins for the Gas Layer on GJ 1214b


\bibitem[Saumon \& Guillot(2004)]{2004ApJ...609.1170S} Saumon, D., \& Guillot, T.\ 2004, \apj, 609, 1170 
%Shock Compression of Deuterium and the Interiors of Jupiter and Saturn

\bibitem[Showman \& Malhotra(1999)]{1999Sci...296...77S} Showman, A.~P., \& Malhotra, R.\ 1999, Science, 296, 77
%The Galilean satellites

\bibitem[Snellen et al.(2008)]{2008A&A...487..357S} Snellen, I.~A.~G., Albrecht, S., de Mooij, E.~J.~W., \& Le Poole, R.~S.\ 2008, \aap, 487, 357
%Ground-based detection of sodium in the transmission spectrum of exoplanet HD 209458b

\bibitem[Sotin et al.(2010)]{2010tfch.book...61S} Sotin, C., Mitri, G., Rappaport, N., Schubert, G., \& Stevenson, D.\ 2010, Titan from Cassini-Huygens, 61 
%Titan's Interior Structure

\bibitem[Stevenson et al.(2010)]{2010Natur.464.1161S} Stevenson, K.~B., Harrington, J., Nymeyer, S., et al.\ 2010, \nat, 464, 1161 
%Possible thermochemical disequilibrium in the atmosphere of the exoplanet GJ 436b

\bibitem[Stevenson \& Lunine(1988)]{1988Icar...75..146S} Stevenson, D.~J., \& Lunine, J.~I.\ 1988, \icarus, 75, 146 
%Rapid formation of Jupiter by diffuse redistribution of water vapor in the solar nebula

\bibitem[Swain et al.(2009)]{2009ApJ...704.1616S} Swain, M.~R., Tinetti, G., Vasisht, G., et al.\ 2009a, \apj, 704, 1616 
%Water, Methane, and Carbon Dioxide Present in the Dayside Spectrum of the Exoplanet HD 209458b

\bibitem[Swain et al.(2009)]{2009ApJ...690L.114S} Swain, M.~R., Vasisht, G., Tinetti, G., et al.\ 2009b, \apjl, 690, L114
%Molecular Signatures in the Near-Infrared Dayside Spectrum of HD 189733b

\bibitem[Swain et al.(2008)]{2008Natur.452..329S} Swain, M.~R., Vasisht, G., \& Tinetti, G.\ 2008, \nat, 452, 329
%The presence of methane in the atmosphere of an extrasolar planet

\bibitem[Thiabaud et al.(2014)]{2014A&A...562A..27T} Thiabaud, A., Marboeuf, U., Alibert, Y., et al.\ 2014, \aap, 562, A27
%From stellar nebula to planets: refractory components

\bibitem[Tinetti et al.(2012)]{2012ExA....34..311T} Tinetti, G., Beaulieu, J.~P., Henning, T., et al.\ 2012, Experimental Astronomy, 34, 311 
%EChO. Exoplanet characterisation observatory

\bibitem[Tinetti \& Griffith(2010)]{2010ASPC..430..115T} Tinetti, G., \& Griffith, C.~A.\ 2010, Pathways Towards Habitable Planets, 430, 115 
%Exploring Extrasolar Worlds Today and Tomorrow

\bibitem[Tinetti et al.(2007)]{2007Natur.448..169T} Tinetti, G., Vidal-Madjar, A., Liang, M.-C., et al.\ 2007, \nat, 448, 169 
%Water vapour in the atmosphere of a transiting extrasolar planet

\bibitem[Udry \& Santos(2007)]{2007ARA&A..45..397U} Udry, S., \& Santos, N.~C.\ 2007, \araa, 45, 397
%Statistical Properties of Exoplanets

\bibitem[Valencia et al.(2013)]{2013ApJ...775...10V} Valencia, D., Guillot, T., Parmentier, V., \& Freedman, R.~S.\ 2013, \apj, 775, 10
%Bulk Composition of GJ 1214b and Other Sub-Neptune Exoplanets

\bibitem[Visser et al.(2011)]{2011A&A...534A.132V} Visser, R., Doty, S.~D., \& van Dishoeck, E.~F.\ 2011, \aap, 534, A132
%The chemical history of molecules in circumstellar disks. II. Gas-phase species


\bibitem[Waldmann et al.(2012)]{2012ApJ...744...35W} Waldmann, I.~P., Tinetti, G., Drossart, P., et al.\ 2012, \apj, 744, 35 
%Ground-based Near-infrared Emission Spectroscopy of HD 189733b

\bibitem[Wong et al.(2004)]{2004Icar..171..153W} Wong, M.~H., Mahaffy, P.~R., Atreya, S.~K., Niemann, H.~B., \& Owen, T.~C.\ 2004, \icarus, 171, 153 
%Updated Galileo probe mass spectrometer measurements of carbon, oxygen, nitrogen, and sulfur on Jupiter

\bibitem[Young(2003)]{2003NewAR..47....1Y} Young, R.~E.\ 2003, \nar, 47, 1 
%The Galileo probe: how it has changed our understanding of Jupiter


\end{thebibliography}
\end{document}